\documentclass[useAMS,usenatbib]{mn2e}
\usepackage[dvips]{graphicx}
\usepackage{amssymb}
\usepackage{longtable}
\newcommand{\dn}{\hbox{$\rm D4000$}}
\newcommand{\hb}{\hbox{H$\beta$}}
\newcommand{\hdg}{\hbox{H$\delta_A$+H$\gamma_A$}}
\newcommand{\mgfep}{\hbox{$\rm [MgFe]^\prime$}}

\newcommand{\mgtwofe}{\hbox{$\rm [Mg_2Fe]$}}
\newcommand{\mgsig}{\hbox{$\rm Mg_2$--$\sigma_V$}}
\newcommand{\mgtwo}{\hbox{$\rm Mg_2$}}
\newcommand{\sigv}{\hbox{$\sigma_V$}}
\newcommand{\mgtfe}{\hbox{$\rm Mgb/\langle Fe\rangle$}}
\newcommand{\rpetro}{\hbox{$\rm R_{50,r}$}}

\def\aj{AJ}%
\def\apj{ApJ}%
\def\apjl{ApJ}%
\def\apjs{ApJS}%
\def\apss{Ap\&SS}%
\def\aap{A\&A}%
\def\mnras{MNRAS}%
\def\pasp{PASP}%
\title[Ages and metallicities of early-type galaxies]
{Ages and metallicities of early-type galaxies in the Sloan Digital Sky
Survey: new insight into the physical origin of the colour-magnitude and
the Mg$_2$--$\sigma_V$ relations}
\author[A. Gallazzi, S. Charlot, J. Brinchmann, S.D.M. White]{Anna 
Gallazzi$^{1}$\thanks{E-mail: gallazzi@MPA-Garching.MPG.DE}, 
St{\'e}phane Charlot$^{1}$~$^,$~$^{2}$, Jarle Brinchmann$^{3}$, Simon D.M.
White$^{1}$\\
$^{1}$Max-Planck-Institut f\"ur Astrophysik, Karl-Schwarzschild-Str. 1, 
D-85748 Garching bei M\"unchen, Germany\\
$^{2}$Institut d'Astrophysique de Paris, UMR7095 CNRS, Universit\'e Pierre \& Marie
     Curie, 98 bis boulevard Arago, 75014 Paris, France \\
$^{3}$Centro de Astrof{\'\i}sica da
Universidade do Porto, Rua das Estrelas - 4150-762 Porto, Portugal}
\begin{document}

\date{Accepted 2006 May 11. Received 2006 May 10; in original form 2005 December 22}

\pagerange{\pageref{firstpage}--\pageref{lastpage}} \pubyear{2006}

\maketitle

  \label{firstpage}

\begin{abstract}
We exploit recent constraints on the ages and metallicities of early-type galaxies in the Sloan
Digital Sky Survey (SDSS) to gain new insight into  the physical origin of two fundamental
relations obeyed by these galaxies: the colour-magnitude and the \mgsig\ relations. Our sample
consists of  26,003 galaxies selected from the SDSS Data Release Two (DR2) on the basis of their
concentrated light profiles, for which we have previously derived median-likelihood estimates of
stellar metallicity, light-weighted age and stellar mass. Our analysis provides the most
unambiguous demonstration to  date of the fact that both the colour-magnitude and the \mgsig\
relations are primarily sequences in stellar mass and that total stellar metallicity,
$\alpha$-elements-to-iron abundance ratio and light-weighted age all increase with mass along the
two relations. For high-mass ellipticals, the dispersion in age is small and consistent with the
error. At the low-mass end, there is a tail towards younger ages, which dominates the scatter in
colour and index strength at fixed mass. A small, but detectable, intrinsic scatter in the
mass-metallicity relation also contributes to the scatter in the two observational scaling
relations, even at high masses. Our results suggest that the chemical composition of an
early-type galaxy is more tightly related to its dynamical mass (including stars and dark matter)
than to its stellar mass. The ratio between stellar mass and dynamical mass appears to decrease
from the least massive to the most massive galaxies in our sample.
\end{abstract}


\section{Introduction}\label{intro}
The observed properties of early-type galaxies obey several fundamental relations, which have
long been thought to hide important clues about the physical processes that influenced the
formation and evolution of these galaxies. For example, luminosity, central velocity dispersion
\sigv, mean surface brightness, brightness profile, colours and \mgtwo\ absorption-line index all
appear to be tightly related to each other in early-type galaxies
\citep{baum59,fj76,vs77,kormendy77,bender93,dd87}. The colour-magnitude and the \mgsig\ relations
are particularly interesting in that they connect the  luminous and dynamical masses of the
galaxies with the physical  properties of their stellar populations. The tightness and
homogeneity of  these two relations must be telling us something fundamental about the  epoch and
the process of formation of early-type galaxies \citep[see for a review][]{renzini06}. 

The physical interpretation of the above observational relations is still a subject of debate.
The colour-magnitude relation is often interpreted as a sequence of increasing metallicity with
increasing luminosity  \citep{faber73,worthey94,kodama97,kodama99}. However, it has been proposed
that, in addition to metallicity, age could at least in part drive the relation
\citep{gonzalez93,terlevich99,fcs99,poggianti01a}.  Studies of evolution with cosmic time
indicate that the slope of the colour-magnitude relation has changed little since $z\sim1$
\citep{kodama97,kodama98,stanford98,blakeslee03}. This has been used as an argument against age
as the primary driver of the relation, since the colours of young stellar  populations evolve
faster than those of old stellar populations.  Similar interpretations have been proposed for the
relation between \mgtwo\ index strength and central velocity dispersion \sigv, which is
generally  thought to arise from a combination of age and metallicity variations 
\citep{colless99,trager00b, kunt01}. 

The difficulty of obtaining unambiguous constraints on the relative  influence of age and
metallicity on the colour-magnitude and the \mgsig\ relations is a consequence of the difficulty
of deriving accurate ages  and metallicities for large samples of early-type galaxies: age and 
metallicity both tend to redden the colours and strengthen the \mgtwo\  absorption line in
similar ways \citep[e.g.,][]{worthey94}. In addition, at fixed metallicity, the
$\alpha$-element-to-iron ($\alpha$/Fe) abundance ratio appears to be larger in the most massive
early-type galaxies than in the nearby stars used to calibrate age and metallicity estimates 
\citep{wfg92,vazdekis01c}. This has been explored recently by  \cite{thomas04}, who used
`closed-box' chemical evolution models with  variable heavy-element abundance ratios to analyse a
heterogeneous sample of 124 nearby early-type galaxies, in both low-density and high-density 
environments. According to these models, massive galaxies formed earlier and more rapidly than
low-mass  galaxies, while both the colour-magnitude and the \mgsig\ relations are primarily
driven by metallicity. \cite{bernardi03a,bernardi03b, bernardi03c,bernardi03d} carried out a more
observationally  oriented analysis on a sample of nearly 9000 early-type galaxies from the Sloan
Digital Sky Survey (SDSS). They showed that the colour-magnitude  relation reflects a dependence
of both colour and luminosity on velocity dispersion. By {\em assuming} that luminosity traces
metallicity and that the scatter in colour at fixed luminosity traces age, \cite{bernardi05}
explored how age and metallicity may be related to velocity dispersion in early-type galaxies.

In this paper, we re-examine the physical origin of the colour-magnitude and the \mgsig\
relations using a different approach. Our starting point is a set of statistical estimates of
light-weighted age, stellar metallicity  and stellar mass for a large sample of 26,003 early-type
galaxies drawn  from the SDSS Data Release Two (DR2). We derived these constraints in  earlier
work by using a comprehensive library of model spectra at  medium-high resolution
\citep[][hereafter Paper~I]{paperI}, to interpret the strengths of five spectral absorption
features with negligible dependence on the $\alpha$/Fe ratio. We use here this dataset, together
with an observational tracer of the $\alpha$/Fe ratio, to demonstrate  unambiguously that both
the colour-magnitude and the \mgsig\ relations of early-type galaxies are primarily sequences in
stellar mass and that both the total stellar metallicity and the $\alpha$/Fe ratio increase with
mass along the two relations. Light-weighted age increases from the least massive to the most
massive early-type galaxies, with a larger spread at low masses that dominates the scatter in the
colour-magnitude and \mgsig\ relations. The small intrinsic scatter in metallicity at fixed mass
also contributes to the scatter in the two scaling relations.

We present our sample in Section~\ref{sample} below, along with a brief  description of the
method adopted in Paper~I to derive statistical  estimates of the ages, metallicities and stellar
masses of the galaxies.  The influence of these parameters on the colour-magnitude and the
\mgsig\  relations is explored in Sections~\ref{cm} and \ref{mgv}, respectively, and their
possible dependence on galaxy environment is addressed in Section~\ref{environment}. In
Section~\ref{mass} we discuss some implications on the relations between physical parameters and
dynamical mass. Section \ref{summary} summarises our conclusions. Throughout the paper we use
$\rm \Omega_m=0.3$, $\rm \Omega_\Lambda=0.7$ and $\rm H_0=70~km~s^{-1}~Mpc^{-1}$

\section{Observational sample}\label{sample}
We select our sample from the main spectroscopic sample of the SDSS DR2  \citep{dr204}. The SDSS
is an imaging and spectroscopic survey of the high Galactic latitude sky, which will obtain $u$,
$g$, $r$, $i$ and $z$  photometry of almost a quarter of the sky and spectra of at least 700,000
objects \citep{York}. The spectra are taken using 3\arcsec-diameter fibres, positioned as close
as possible to the centres of the target galaxies.  Stellar metallicity, light-weighted age and
stellar mass estimates are available  from Paper~I for a sample of 175,128 galaxies with
Petrosian $r$-band  magnitudes in the range $14.5<r<17.77$ (after correction for foreground
Galactic extinction) and in the redshift range $0.005<z\leq0.22$.  The lower redshift limit
allows us to include low-luminosity galaxies (corresponding to a stellar mass of $\sim 10^8
M_\odot$), while still avoiding  redshifts for which deviations from the Hubble flow can be
substantial. The upper limit corresponds roughly to the redshift at which a typical 
$10^{11}M_\odot$ galaxy is detected with median S/N per pixel greater than 20.

We select early-type galaxies on the basis of the light concentration  index $C=R_{90}/R_{50}$,
defined as the ratio of the radii enclosing 90  and 50 percent of the total Petrosian $r$-band
luminosity of a galaxy. This parameter has been shown to correlate well with morphological type
\citep{strateva01,shimasaku01}. Thus, it allows a rough morphological  classification of SDSS
galaxies. \cite{strateva01} propose a cut at $C=2.6$ to separate early- from late-type galaxies.
To limit the contamination  by disc galaxies with large bulges, we define here as `early-type'
those  galaxies with concentration index $C\geq2.8$. In this way, we select 67,411 early-type
galaxies in the redshift range $0.005<z\leq0.22$. We note that we decided against further
limiting the contamination of our sample by systems with residual star formation by imposing a
lower cutoff in 4000-\AA\ break strength or an upper limit on emission lines equivalent width.
Since these quantities correlate with colour, such limits would introduce an unwanted cutoff in
the colour-magnitude  relation.\footnote{\cite{bernardi05} adopted a different criterion to 
select early-type galaxies from the SDSS. They considered to be of early type those galaxies for
which the $r$-band surface brightness profile is better described by a de~Vaucouleurs law than by
an exponential law  (photometric parameter ${\tt fracDev}> 0.8$) and that do not have emission
lines (spectroscopic parameter ${\tt eclass}<0$).}

Bayesian-likelihood estimates of the $r$-band light-weighted ages, stellar  metallicities and
stellar masses of the 67,411 galaxies in our sample are available from Paper~I. These estimates
were derived by comparing the  spectrum of each galaxy to a library of \cite{bc03} models at
medium-high spectral resolution, encompassing the full range of  physically plausible star
formation histories. In practice, we compared the strengths of five spectral absorption features
in the spectrum of each  observed galaxy to the strengths of these features in every model
spectrum (broadened to the observed velocity dispersion) in the library. We used  \dn, \hb\ and
\hdg\ as age-sensitive indices and  \mgtwofe\ and \mgfep\ as metal-sensitive indices, all of
which depend negligibly on the $\alpha$/Fe ratio.\footnote{This might not be entirely correct for
H$\delta_A$ and H$\gamma_A$, which are suspected to depend on the $\alpha$/Fe ratio at high
metallicity \citep{thomas04,korn05}. However, we did not find any discrepancy between the
metallicities and ages derived including or excluding \hdg\ (see section 2.4.2 of Paper~I).} This
comparison allowed us to construct the probability density functions of age, metallicity and
stellar mass for every galaxy.  The estimate of each parameter is given by the median of the
corresponding probability distribution, while the $\pm1\sigma$ error on each parameter is given
by half the $16-84$ percent percentile range of the likelihood distribution (this would be
equivalent to the $\pm1\sigma$ range for a Gaussian distribution).

We note that the stellar ages, as well as the other physical parameters, are derived by fitting
the galaxy spectra as observed and so refer to the galaxies at the time they are observed.
Because of the bright and faint magnitude limits in our sample, there is a strong correlation
between luminosity and redshift. If uncorrected, it may introduce systematic effects in
correlations between age and luminosity (or mass). To avoid this, we correct our measured ages so
that they are relative to the present, rather than to the point of observation by adding to the
measured age for each galaxy the look-back time to the redshift at which it is observed. The
metallicity is left unchanged. This look-back time varies from 0.07 to 2.64 Gyr over the redshift
range covered by our sample. The corrections mainly affect the most luminous galaxies, which are
found out to higher redshifts. All  light-weighted ages quoted throughout this paper refer to
$z=0$.

As described in Paper~I, our constraints on metallicity and age are  sensitive to the
observational signal-to-noise ratio (S/N) of the spectra. A median S/N per pixel of at least 20
is required to constrain metallicity  reliably. For this reason, we consider here only those
galaxies with a  median S/N per pixel greater than 20. This cut reduces our sample to 26,003
high-concentration galaxies in the redshift range $0.005<z\leq0.22$. This is the same sample of
`early-type' galaxies as analysed in section 3.3 of Paper~I. As shown there, the S/N requirement 
biases the sample towards low-redshift, high-surface brightness galaxies, but it does not
introduce any bias in the luminosity, colour, velocity-dispersion and  index-strength
distributions of the early-type galaxy sample.

Our selection of early-type galaxies, based only on the concentration parameter, includes
early-type spiral galaxies, galaxies with emission lines and  active galactic nuclei (AGN). We
can divide our sample into five subclasses according to the classification given by
\cite{Jarle03} on the basis of the emission-line properties of SDSS-DR2 galaxies. Following their
notation, we label as (1)  unclassifiable (`Unclass.') those galaxies which cannot be classified
using the \cite{BPT} diagram, i.e. mostly galaxies with no or weak emission lines; (2) AGN those
galaxies with a substantial AGN contribution to their emission-line fluxes; (3) star forming
(`SF') those galaxies with S/N$>3$ in H$\beta$, H$\alpha$, [OIII]5007, [NII]6564 emission lines
and for which the contribution to H$\alpha$ from AGN is less than 1 percent; (4) composite (`C')
those galaxies with S/N$>3$ in the same four lines, but for which the AGN contribution to
H$\alpha$ luminosity can be up to 40 percent; (5) low-S/N star forming (`low-S/N SF') those
galaxies with S/N$>2$ in H$\alpha$ but without AGN contribution to their spectra. The sample
includes 10,982 unclassifiable galaxies, 7782 AGN and 3018 low-S/N SF galaxies. There are 2858
composite and 1362 star-forming galaxies, accounting for roughly 11 and 5 percent of the total
sample, respectively. As expected, most galaxies in the sample have  specific star formation
rates (SFR/M$_\ast$) less than $10^{-10.6}$ yr$^{-1}$, which  are in most cases consistent with
zero. We note that the distribution in specific star formation rates for the subclasses of C and
SF galaxies shows a peak at  $10^{-10.0}$ yr$^{-1}$.

\begin{figure*}
\centerline{\includegraphics[width=10truecm]{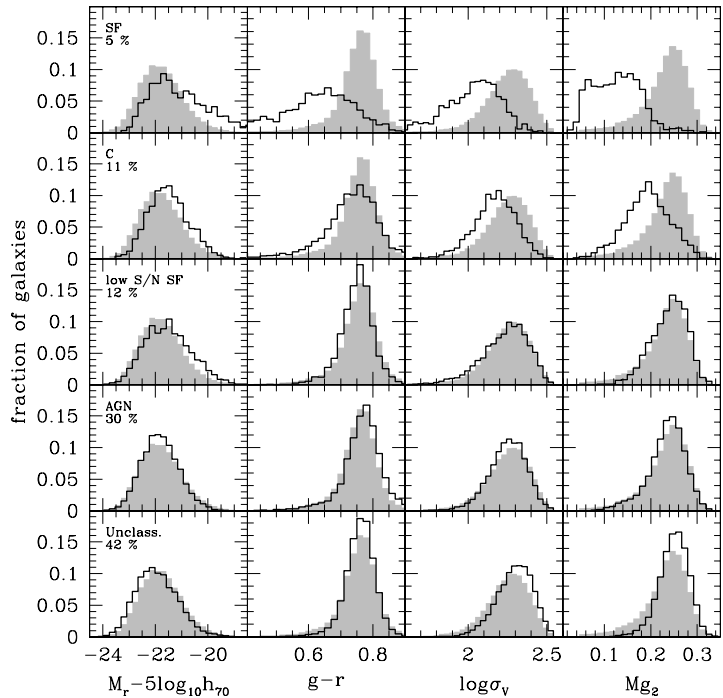}}
\caption{Distributions of the $r$-band absolute magnitude, $g-r$ colour  ({\it k}-corrected to
$z=0$ and corrected for evolution following Bernardi et al. 2005), logarithm of velocity
dispersion (in $\rm km~s^{-1}$)  and Mg$_2$ index strength (in mag) for a sample of 26,003
early-type (high-concentration)  SDSS-DR2 galaxies with median S/N  per pixel greater than 20
(grey histograms,  repeated in all rows). From top to bottom, solid histograms in each row show
the  distributions for subsamples of galaxies classified by \citet{Jarle03} according  to their
emission-line properties as star-forming (`SF'), composite (`C'), low  signal-to-noise
star-forming (`low-S/N SF'), AGN (`AGN') and unclassifiable  (`Unclass.', these are mainly
galaxies with no emission lines). The fractional contributions by the different classes to the
total sample of 26,003 early-type galaxies are indicated in the left-most panels. All histograms
are normalised to unit area.}\label{sample_distr} 
\end{figure*}

Fig.~\ref{sample_distr} shows the distributions of several observational quantities  of interest
to us for this sample: the $r$-band absolute magnitude $M_r$, the $g-r$ colour, the logarithm of
the velocity dispersion (in  $\rm km~s^{-1}$) and the Mg$_2$ index strength. The magnitude $M_r$
and the colour $g-r$ represent rest-frame quantities. They are corrected for evolution following
the prescription of \cite{bernardi05}, as described in Section~\ref{obscm}. Also the \mgtwo\
index strength is corrected to $z=0$ assuming passive evolution, as described in
Section~\ref{mgv}. In each row, the solid histogram represents the distribution for each of the
above subclasses (whose fractional contribution to the total sample is given in the leftmost
panel), while the distribution for the sample as a whole is represented by the shaded grey
histogram. The distributions for the unclassifiable, AGN and low-S/N SF galaxies agree well with
the distributions for the sample as a whole. In contrast, C  and especially SF galaxies tend to
concentrate in the low-luminosity, blue,  low-velocity-dispersion and low-Mg$_2$ tails of the
distributions. 

\begin{figure}
\centerline{\includegraphics[width=9truecm]{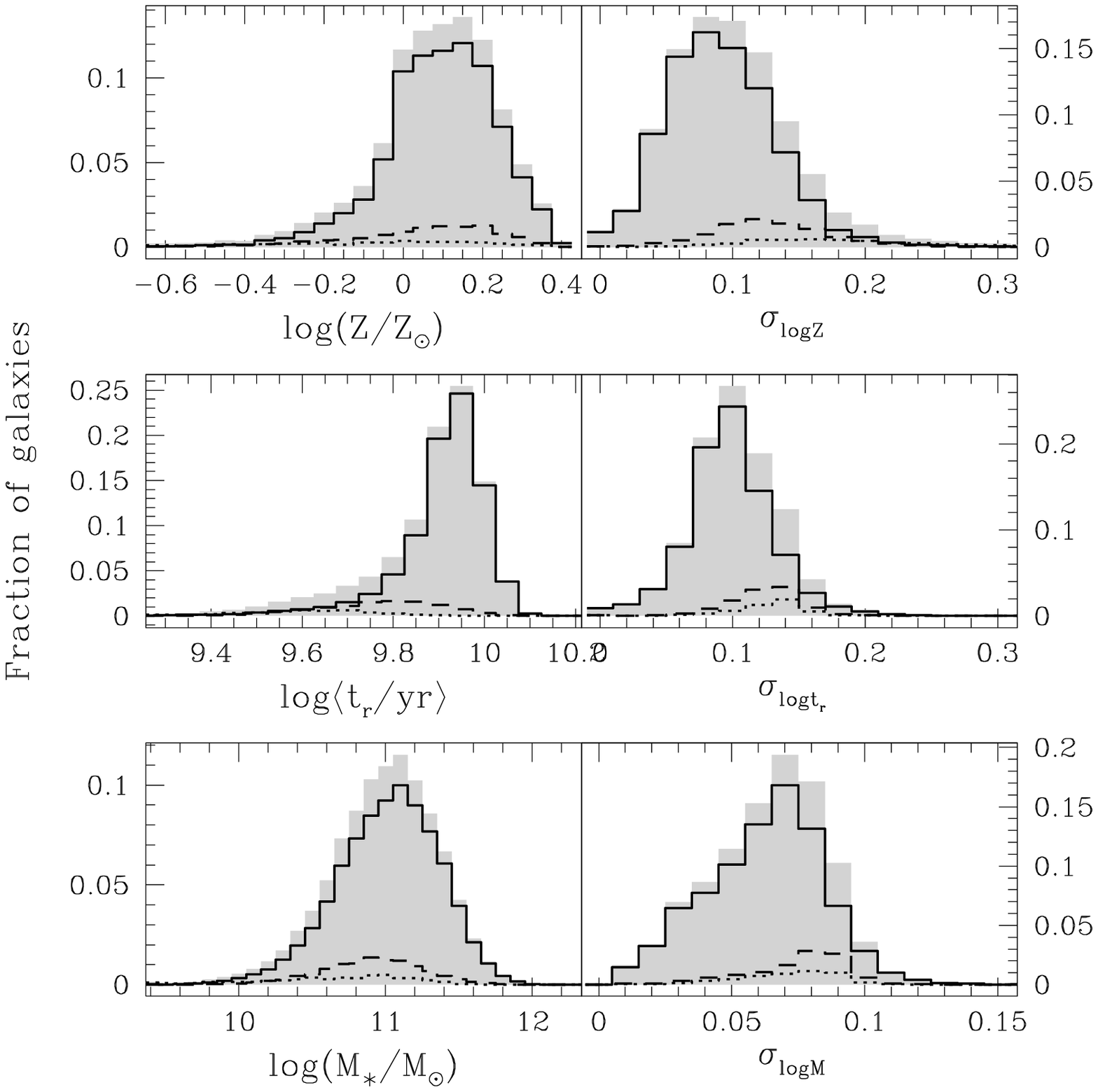}}
\caption{Left: distributions of the median-likelihood estimates of stellar metallicity (top),
$r$-band light-weighted age (middle) and stellar mass (bottom) for the same sample of early-type
galaxies as in Fig.~\ref{sample_distr}. The dotted histogram shows the distribution for SF
galaxies, the dashed histogram for C galaxies and the solid histogram for the rest of the sample
(Unclass, AGN and  low-S/N SF). Right: distributions of the associated errors, computed as one
half the 68 percent confidence ranges in the estimates of $\log(Z/Z_\odot)$, $\log(t_r/{\rm 
yr})$ and $\log(M_\ast/M_\odot)$.}
\label{ztm_distr}
\end{figure}

The stellar metallicities, ages and stellar masses derived in Paper~I for the galaxies in the
sample are shown in Fig.~\ref{ztm_distr} (shaded grey histograms). The distribution in 
metallicity extends from $0.4\,Z_\odot$ to $2\,Z_\odot$ with a peak  around $Z=1.6\,Z_\odot$. The
distribution in $r$-band light-weighted age  extends from 2.5 to 12~Gyr with a peak around
$t_r=9$~Gyr. The distribution in stellar mass extends from $10^{10}$ to $10^{12}\,M_\odot$ with a
peak around $M_\ast=10^{11}\,M_\odot$. The right-hand panels of Fig.~\ref{ztm_distr} show the
distributions of the associated errors, computed as one half the  68 percent confidence ranges in
the estimates of $\log(Z/Z_\odot)$,  $\log(t_r/{\rm yr})$ and $\log (M_\ast/M_\odot)$. The
typical uncertainty in both metallicity and age estimates is about $\pm0.1$ dex, while stellar
mass is constrained to better than $\pm0.1$ dex for the majority of the galaxies in the sample.
The same is true when considering Unclass., AGN and low-S/N SF galaxies only (solid histograms).
The distributions for SF (dotted line) and C (dashed line) galaxies extend to lower stellar
metallicities, ages and stellar masses than the bulk of the sample, with slightly higher errors
in all three parameters. 

In addition to the effects of age, metallicity and stellar mass, we are also interested in the
influence of the $\alpha$/Fe ratio on the observed  properties of early-type galaxies. This ratio
can be empirically quantified by the relative strengths of Mg- and Fe-based absorption-line
indices  \citep{Thomas03a}. We use here \mgtfe, where $\rm \langle Fe\rangle$ is the average of
the Fe5270 and  Fe5335 index strengths. In what follows, we compare the observed \mgtfe\ ratio 
of a galaxy to that of the model that best reproduces the 5 spectral features mentioned above.
Since the models have scaled-solar abundance ratios, any discrepancy $\rm \Delta(\mgtfe)$ between
observed and model index strengths  can then be interpreted as a departure of the
$\alpha$-elements-to-iron abundance ratio from solar in the observed galaxy. 

Working with the difference in index ratio rather than with the observed value of \mgtfe\ allows
us to take into account the dependence of the index strength on age. It is also interesting to
check whether the difference $\rm \Delta(\mgtfe)$ traces linearly the true $\alpha$/Fe abundance
ratio over the whole parameter space covered by our galaxies. It is not possible to check this
self-consistently on our sample with our own models, since they do not include variations in
element abundance ratios. For this test, we used Simple Stellar Population (SSP) models with
variable element abundance ratios from \cite{Thomas03a}. We have considered SSPs with different
$\alpha$/Fe abundance ratios ([$\alpha$/Fe]=$-0.2,0,0.2,0.4,0.6$), with age varying from 1 to 13
Gyr and metallicity varying from $\log(Z/Z_\odot)=-0.6$ to 0.5. For each model we calculate the
difference in \mgtfe\ with respect to the corresponding scaled-solar model at the same age and
metallicity. We have looked at the relation between $\rm \Delta(\mgtfe)$ and the abundance ratio
[$\alpha$/Fe] at fixed metallicity. We do not find significant differences in the slope and
zero-point of the linear relations, averaged over age, for different metallicities. Only at
$\log(Z/Z_\odot)\sim-0.5$ is the relation somewhat flatter; we note that this is not relevant for
the galaxies in our sample, which predominantly have metallicities above $\log(Z/Z_\odot)=-0.2$
(see Fig.~\ref{ztm_distr}). The proportionality between $\rm \Delta(\mgtfe)$ and [$\alpha$/Fe] is
confirmed by plotting $\rm \Delta(\mgtfe)$ as a function of age and as a function of metallicity
for SSPs with different $\alpha$/Fe ratio. At fixed [$\alpha$/Fe], the measured $\rm
\Delta(\mgtfe)$ is constant over the age and metallicity ranges covered by our sample. These
checks reassure us that the discrepancy $\rm \Delta(\mgtfe)$ between the observed and the
best-fit model index strengths has the same proportionality with [$\alpha$/Fe] over the age and
metallicity ranges covered by the sample. 

\section{Physical origin of observed relations for early-type galaxies}
\label{ET}
We use here the sample of 26,003 early-type galaxies described above to investigate the physical 
origin of two fundamental relations obeyed by early-type galaxies: the  colour-magnitude relation
(Section~\ref{cm}) and the \mgsig\ relation (Section~\ref{mgv}). We address the environmental
dependence of these relations and of galaxy physical parameters in Section~\ref{environment}. We
then explore in more detail how physical parameters are related to the depth of a galaxy's
potential well (Section~\ref{mass}).

\subsection{The colour-magnitude relation}\label{cm} 
The optical colours of early-type galaxies are strongly correlated with  absolute magnitude, in
the sense that bright galaxies tend to be redder  than faint galaxies
\citep[e.g.][]{baum59,deV61,faber73,vs77,ble92}. As mentioned in Section~\ref{intro}, the
physical origin of this correlation is still a matter of controversy, mainly because of the lack
of large  samples of galaxies with accurate age and metallicity estimates.

\subsubsection{Observed colour-magnitude relation}\label{obscm}
We consider here the colour-magnitude relation (CMR) defined by the rest-frame $g-r$ colour and
the absolute $r$-band magnitude $M_r$. We compute these quantities using the SDSS {\tt model}
magnitudes, which are obtained from fits of the $r$-band surface brightness profile of each
galaxy with either a de~Vaucouleurs law or an exponential law (the best-fitting profile is  also
adopted in the other photometric bands). We {\it k}-correct the  observed $g-r$ colour and the
$M_r$ absolute magnitude to $z=0$ using the model in the spectral library of Section~\ref{sample}
that best reproduces the spectral absorption features of the galaxy, reddened to reproduce the
observed colours at the redshift of the galaxy (this reddening correction is typically small; see
below). We also correct colours and magnitudes for evolution, adopting the correction estimated
by \cite{bernardi05} from a sample of 39,320 SDSS early-type galaxies, i.e. we make the
magnitudes fainter by $0.85z$ and the colours redder by $0.3z$. From now on, we  denote by $M_r$
and $g-r$ the {\it k}+{\it e}-corrected (but not dereddened) quantities.

Fig.~\ref{CM_dens} shows the $g-r$, $M_r$ colour-magnitude relation defined in this  way  by
early-type galaxies in the sample described in Section~\ref{sample}. We have arranged galaxies in
bins of absolute magnitude and colour (0.15 and 0.01 width  respectively). The grey scale
indicates, for each magnitude bin, the relative  distribution of galaxies in the different colour
bins (we do not show bins  containing less than 2 percent of the total number of galaxies at a
given magnitude).  The straight line in Fig.~\ref{CM_dens} shows a `robust' fit of these data,
obtained by minimising the absolute deviation in colour as a function of magnitude. We note that
this lies slightly to the blue of the ridge-line of the relation, reflecting the skewed colour 
distributions of Fig.~\ref{sample_distr}. The robust fit has a slope\footnote{The errors on the
slopes of the fitted relations quoted throughout the paper have been estimated using a
`jackknife' method. The parameters of the fitted relations are summarised in Table~\ref{fits}.}
of $-0.024\pm0.002$. This slope is consistent with that of $-0.024$  obtained previously by
\cite{bernardi05}\footnote{We estimated this slope as $\xi_{CM}  \sigma_{CC}/\sigma_{MM}$, using
the values reported in tables 1 and 2 of \citet[see their appendix B1]{bernardi05}.}. The dashed
lines in  Fig.~\ref{CM_dens} show the mean positive and negative absolute deviations in colour
($\pm 0.050$) relative to this fit.

To better understand the influence of our galaxy selection criteria on our results, we
constructed a second sample (we call this `sample B') following the prescription by Bernardi et
al.. We first select all objects classified as galaxies, with measured velocity dispersions and
with Petrosian $r$-band magnitudes (corrected for galactic reddening) between 14.5 and 17.75 in
the SDSS DR2. We then define as early-type those galaxies with ${\tt fracDev\_r}>0.8$ and ${\tt 
eclass}<0$. We select in this way 59,907 early-type galaxies, for which we have  $g-r$ colours
and $M_r$ absolute magnitudes.\footnote{Note that the sample used by \citet{bernardi05} is
slightly smaller than the SDSS DR2. Note also that sample~B is larger than our primary sample,
mainly because of our cut in S/N ratio.}

In Fig.~\ref{bernardi_comp}, we compare the CMRs obtained from our primary sample and  from
sample~B in different redshift bins \citep[by analogy with fig.~2 of][]{bernardi05}. The black
dot-dashed lines in Fig.~\ref{bernardi_comp} show robust fits to the relations for the two
samples. The slope for sample B ($-0.028 \pm 0.001$, upper panel) is in good agreement with that
obtained by \citet[black solid line] {bernardi05} and with that derived for our primary sample
($-0.024\pm0.002$, lower panel). The colour-magnitude distributions of the galaxies in different
redshift bins in sample B (upper panel) also agree reasonably well with  those in fig.~2 of
\cite{bernardi05}. Thus, our results agree with those of \cite{bernardi05} under similar
assumptions. 

We note that our primary sample differs from sample~B mainly through the inclusion of blue
galaxies at intermediate-to-low luminosities. These are predominantly galaxies in the SF and C
classes, which are likely to be bulge-dominated spiral galaxies with  detectable emission lines
({\tt eclass>0}, which excludes them from sample B). The  contamination of our primary sample by
these galaxies also results from the fact that we do not make any distinction on the basis of
galaxy environment. While the slope of the CMR in low-density environments is not significantly
different from that found for cluster early-type galaxies, the scatter about the relation appears
to increase from high- to low-density environments \citep{ltc80,hogg04}. Also, at the median 
redshift $z\sim0.1$ of our sample, \cite{pimbblet02} find that, at fixed luminosity,  early-type
galaxies tend to have bluer colours in the outskirts than in the centres of galaxy clusters
\citep[see also][]{abraham96}. They interpret this as a difference in age of the galaxies in
different environments.

As in other work on early-type galaxies \citep[e.g.][]{bernardi05}, we do not include any
correction for dust attenuation in the CMR presented here. We can estimate the  influence of dust
on the CMR, using the model spectral fits described in  Section~\ref{sample}. The $g$- and
$r$-band dust attenuation can be estimated from the difference between the emission-line
corrected $r-i$ {\tt fibre}{\footnote{{\tt fibre} magnitudes are obtained from the flux measured
within an aperture of radius 1.5 arcsec, i.e. equal to the fibre radius.} colour of the galaxy
and the $r-i$ colour of the  redshifted, dust-free model, assuming a single power-law attenuation
curve (see also  Paper~I). To quantify the effect of dust attenuation on the CMR, we show in 
Fig.~\ref{CM_dust} the average attenuation vectors (red arrows) for galaxies  falling  into
different bins of $g-r$ and $M_r$ (dashed boxes). As expected, the average  attenuation is
reduced if C and SF galaxies are excluded (blue arrows). In both cases, attenuation by dust
becomes more important at lower luminosities.  Correcting for dust attenuation would thus steepen
the relation. When de-reddening the colours and magnitudes of the galaxies in the sample
according to the average attenuation vectors shown in Fig.~\ref{CM_dust}, we derive a slope of
$-0.035\pm0.002$. Table~\ref{fits} summarizes the parameters of the fits to the ({\it
k}-corrected)  colour-magnitude relation, with and without evolution and dust corrections.

\begin{figure}
\centerline{\includegraphics[width=8truecm]{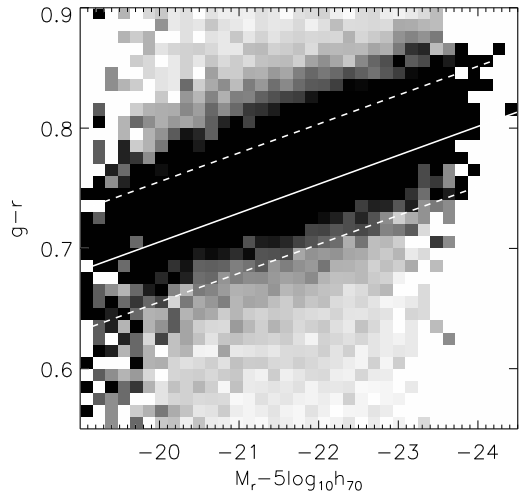}}
\caption{Relation between ({\it k}+{\it e}-corrected) $g-r$ colour and $r$-band absolute
magnitude for the same sample of early-type galaxies as in Fig.~\ref{sample_distr}. The grey
scale indicates, for each magnitude bin, the relative distribution of  galaxies in the different
colour bins (normalised along the colour axis). The solid line represents the relation between
$g-r$ and $M_r$, fitted as described in the text.  The dashed lines show the vertical scatter
about the relation.}\label{CM_dens}
\end{figure} 
\begin{figure}
\centerline{\includegraphics[width=8truecm]{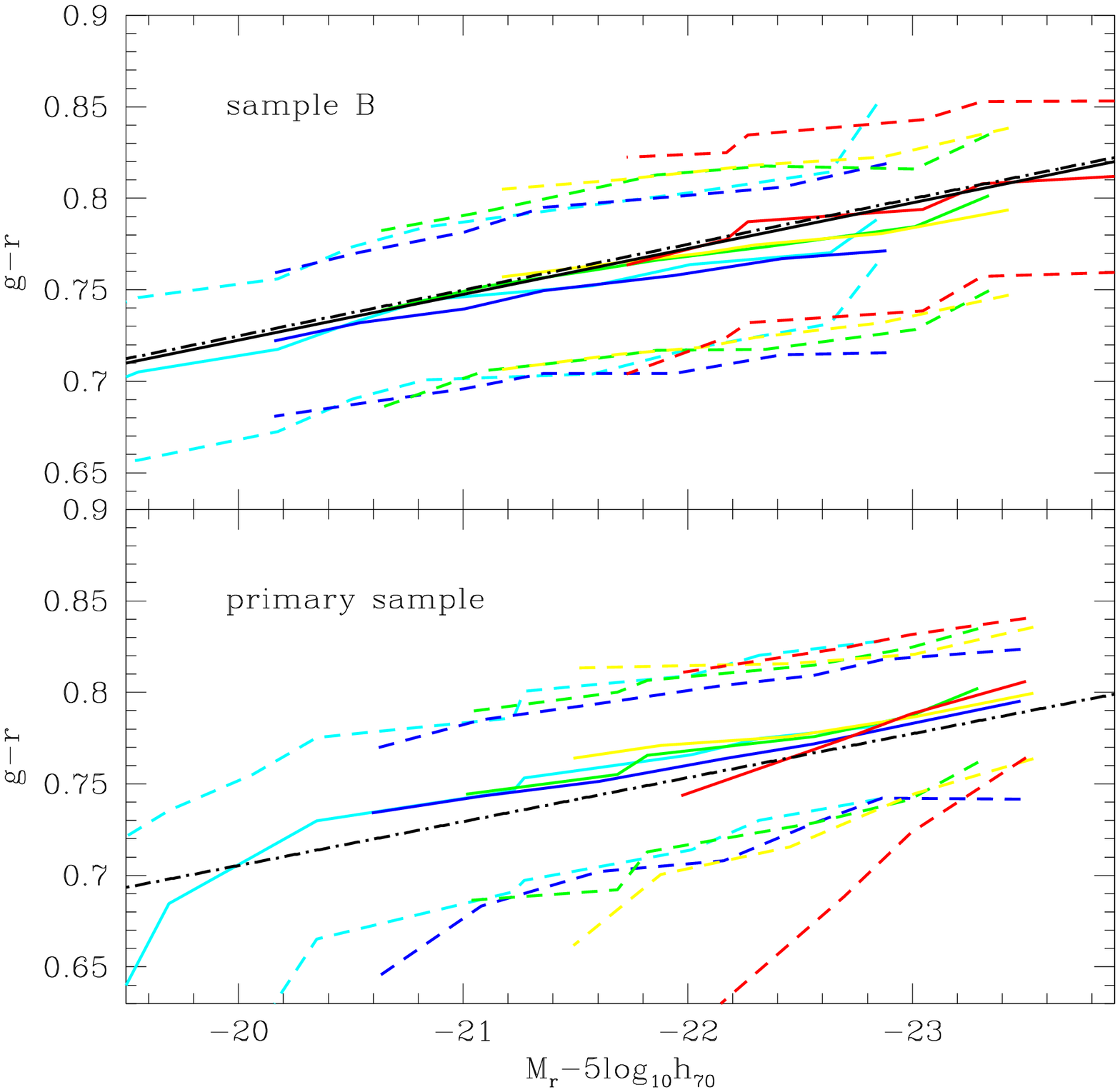}}
\caption{Colour-magnitude relation for the same sample of early-type galaxies as in
Fig.~\ref{sample_distr} (lower panel) and for a sample obtained using the same  selection
criteria as in \citet{bernardi05} (upper panel).  The colours and luminosities of the galaxies in
both samples are corrected for evolution following \citet[i.e., making $g-r$ redder by 0.3$z$ and
$M_r$ fainter by 0.85$z$]{bernardi05}. Different colours refer to different redshift intervals,
$0.02<z<0.07$ (cyan),  $0.07<z<0.09$ (blue), $0.09<z<0.12$ (green), $0.12<z<0.15$ (yellow),
$0.15<z<0.2$ (red). For each redshift bin, solid and dashed lines give the median and the 68 
percent range in colour as a function of luminosity. Black dot-dashed lines show  robust fits to
the relations for the two samples (including galaxies at all redshifts), while the black solid
line in the upper panel is the corresponding relation fitted by \citet{bernardi05} with a
maximum-likelihood technique (see their fig.~2).}
\label{bernardi_comp}
\end{figure}

\begin{figure}
\centerline{\includegraphics[width=6truecm]{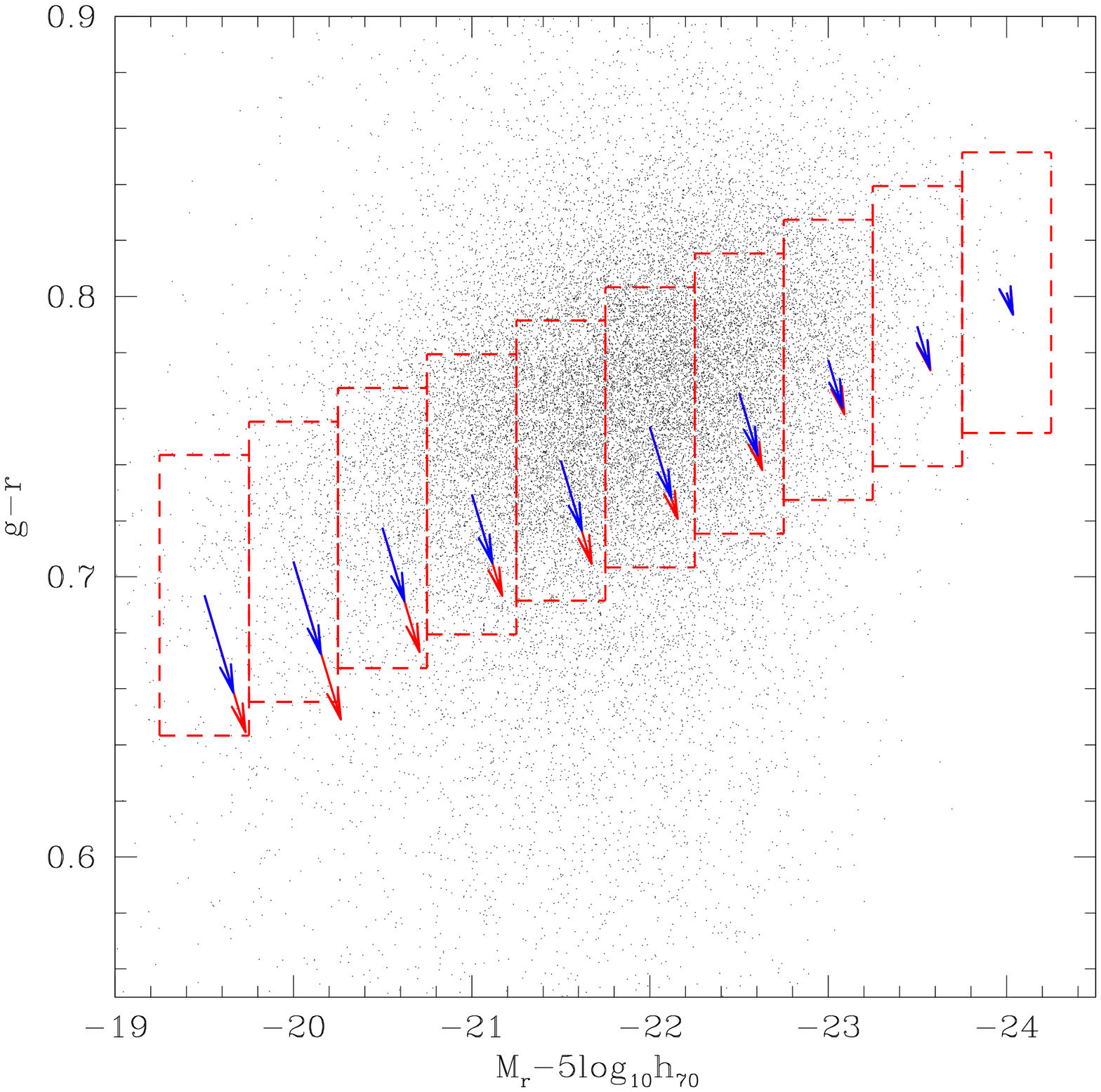}}
\caption{Relation between ({\it k}+{\it e}-corrected) $g-r$ colour and $r$-band absolute
magnitude for the same sample of early-type galaxies as in Fig.~\ref{sample_distr}. The red
arrows indicate the average correction for dust attenuation on the colour  and absolute magnitude
of the galaxies falling into different $g-r,M_r$ bins  (indicated by the dashed boxes) along the
fitted relation. The blue arrows indicate the average dust correction when galaxies in the
star-forming and composite classes are excluded.}\label{CM_dust}
\end{figure} 

\subsubsection{Physical origin of the colour-magnitude relation}\label{physcm}

Fig.~\ref{CM_code} shows how metallicity, age, $\alpha$/Fe ratio and stellar mass  change along
the colour-magnitude relation. We have binned the $g-r$, $M_r$ plane  in the same way as in Fig.
\ref{CM_dens}, the brightness of each $g-r$ bin at a given $M_r$ being proportional to the number
of galaxies falling into this bin (see Section~\ref{obscm}). In each panel, the colour code
reflects the average  properties of the galaxies falling into each colour-magnitude bin, as
indicated.  Panels (a)--(c) show how (the median-likelihood estimates of) age, metallicity and 
stellar mass are distributed along the CMR. Panel (d) shows the distribution of the offset $\rm
\Delta(\mgtfe)$ between observed and predicted \mgtfe\ index strengths, which traces the
$\alpha$-elements-to-iron abundance ratio (Section~\ref{sample}).

\begin{figure*}
\centerline{\includegraphics[width=12truecm]{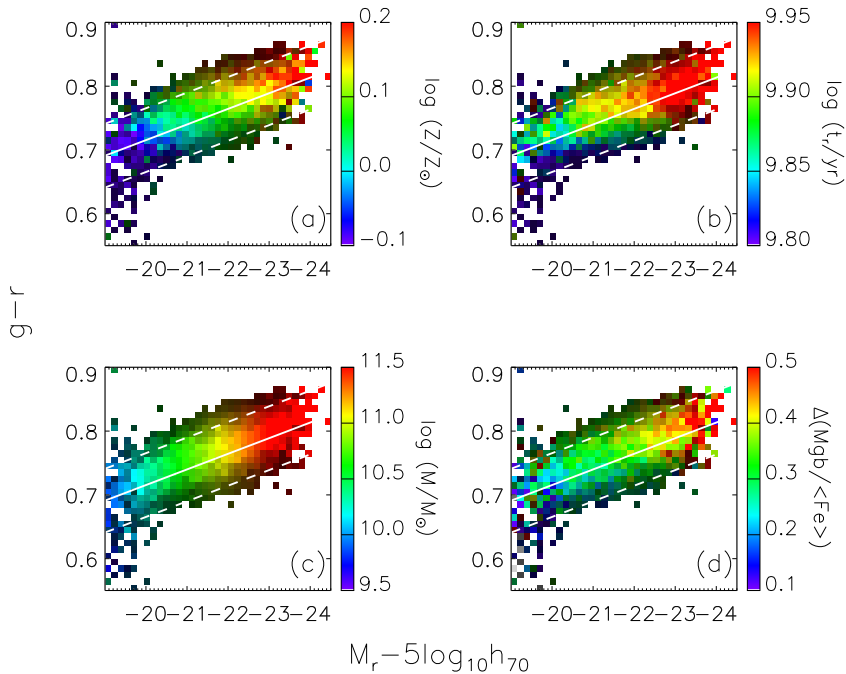}}
\caption{Relation between ({\it k}+{\it e}-corrected) $g-r$ colour and $M_r$ absolute magnitude
for the same sample of early-type galaxies as in Fig.~\ref{sample_distr}. The colour-magnitude
relation  has been binned and colour-coded to reflect the average stellar metallicity (panel a),
$r$-band light-weighted age (panel b), stellar mass (panel c) and the $\alpha$/Fe-indicator
$\Delta(\mgtfe)$ (panel d) of the galaxies falling  into each bin. The shading indicates the
fraction of galaxies populating each colour bin at fixed magnitude, according to the map of Fig.
\ref{CM_dens}.}\label{CM_code}
\end{figure*} 

The most remarkable result from Fig.~\ref{CM_code} is that the colour-magnitude relation is
primarily a sequence in the stellar mass of early-type galaxies (panel c; note that the gradient
in stellar mass is not exactly parallel to the magnitude axis, reflecting a non-constant
mass-to-light ratio in the optical bands). Another striking result is that the average
metallicity increases from $\sim0.8 Z_\odot$ to $\sim1.6 Z_\odot$ from the  faintest to the
brightest galaxies along the relation, the average light-weighted age increasing  by less than 3
Gyrs (from 6.5 and 9.0~Gyr) over the same interval of 5 magnitudes in $M_r$  (panels a and b).
The marked increase in metallicity is accompanied by an increase  in the
$\alpha$-elements-to-iron abundance ratio (panel~d). This is consistent with the finding that
giant elliptical galaxies have higher [Mg/Fe] than faint elliptical  galaxies
\citep[e.g,][]{wfg92,trager00b,kunt01,thomas04}. The gradients in both metallicity and age are
not exactly parallel to the relation, but at fixed magnitude, bluer galaxies are more metal-poor
and younger than redder galaxies. Thus, both metallicity and age contribute to the scatter about
the relation.

It is of interest to compare the relative contributions by age and metallicity to the scatter
about the CMR. We note that we expect dust attenuation and changes in the  [$\alpha$/Fe] ratio to
contribute negligibly to the scatter, since the scatter in  $g-r$ colour of dust-free,
[$\alpha$/Fe]=0 models with the ages and metallicities of  the galaxies in our sample
($\pm0.041$)\footnote{This is estimated by adopting, for  each galaxy in the sample, the median
of the likelihood distribution in $g-r$ obtained by weighting each model in the library of
Section~\ref{sample} by $\chi^2=[(\log Z_{mod} -\log Z)/\sigma_{\log Z}]^2+[(\log t_{mod}-\log
t_r)/\sigma_{\log t_r}]^2$.} is very similar to the observed scatter ($\pm0.050$). For
comparison, the scatter in $g-r$ colour of models with the metallicities of the galaxies in our
sample but with a fixed imposed (average) age at fixed $M_r$ is only $\pm0.022$, while that of
models with the ages of the galaxies in our sample but with a fixed imposed (average) metallicity
at fixed $M_r$ is $\pm0.028$. This indicates that age and metallicity contribute a similar amount
to the scatter in the  colour-magnitude relation for early-type galaxies.

Figure~\ref{CM_spread} better quantifies the intrinsic scatter in the two physical parameters and
how much it determines the scatter in colour as a function of stellar mass. The left-hand panels
show the distribution in light-weighted age (or metallicity) as a function of $g-r$ colour in six
bins of stellar mass (the median and the $16-84$ interpercentile range are given by the red
points and associated error bars). The right-hand panels show the marginalized distribution in
age (metallicity) compared with a Gaussian of width equal to the average error in age
(metallicity) for the galaxies in each stellar mass bin (dotted curve). From the left-hand plot
we can see that at masses above $10^{11}M_\odot$ the scatter in light-weighted age can be
entirely accounted for by the measurement errors, which are on average less than 0.1~dex. We can
thus say that all massive ellipticals of given mass have the same mean age (within the errors). 
At lower masses the scatter in light-weighted age becomes larger than the typical error, and the
distribution is skewed toward younger ages. At fixed mass, there is a correlation between age and
colour, which saturates for the reddest galaxies. We also detect a small intrinsic scatter in the
mass-metallicity relation in all mass bins (right-hand plot), which contributes to the scatter in
the CMR at fixed mass, even at the high-mass end.

\begin{figure*}
\centerline{\includegraphics[width=8truecm]{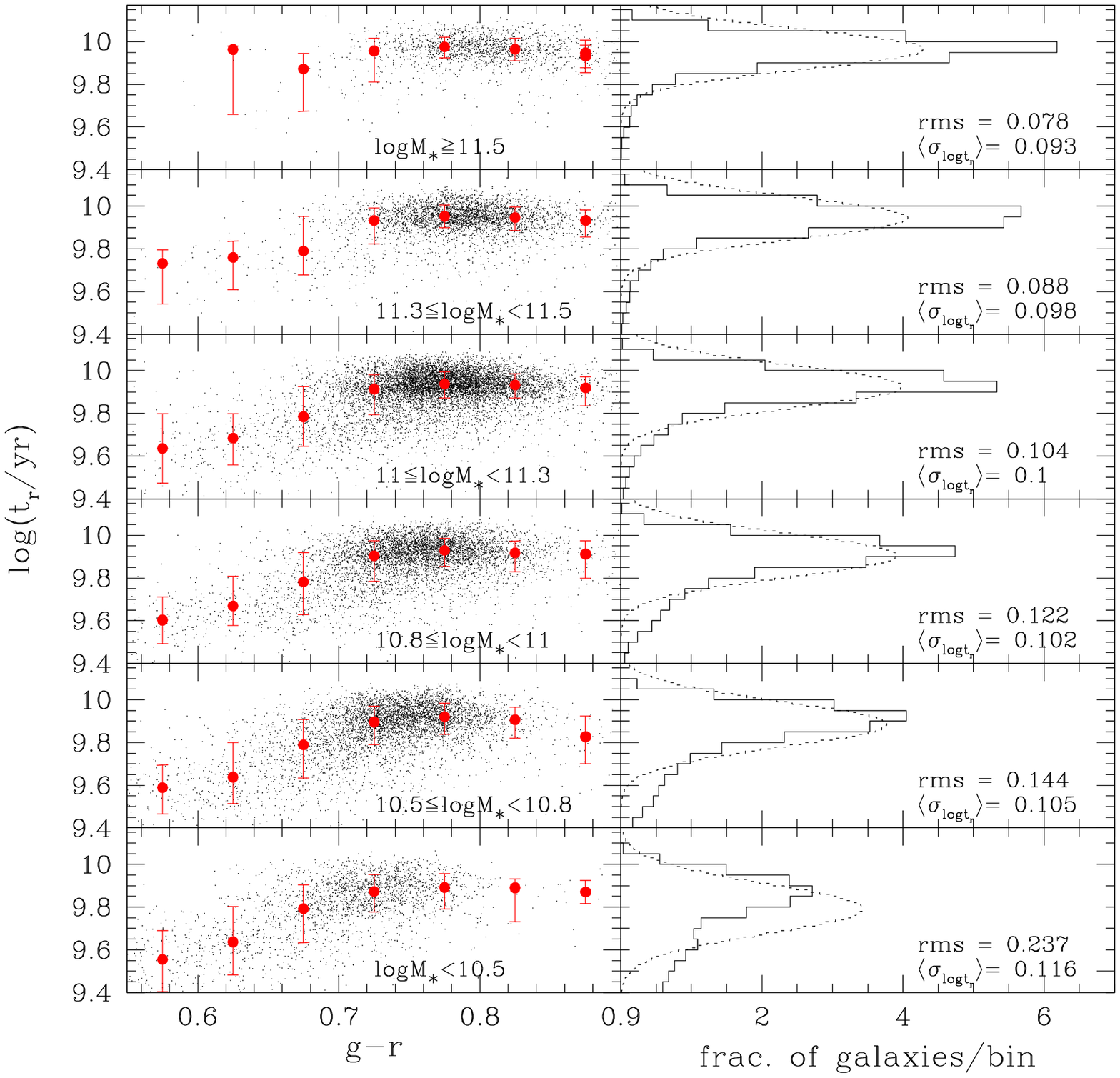}
\includegraphics[width=8truecm]{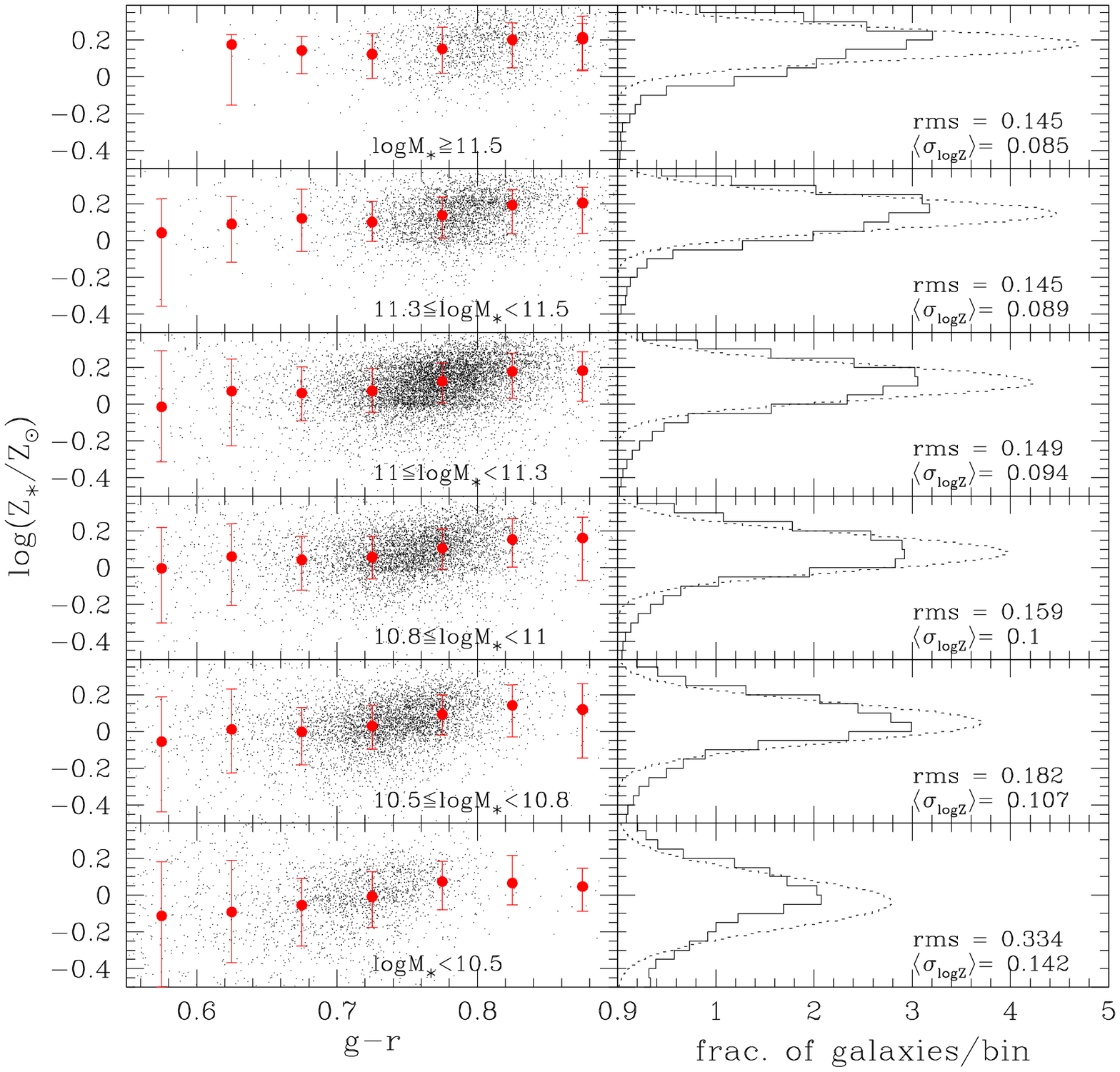}}
\caption{Left panels: distribution in light-weighted age (left plot) and stellar metallicity
(right plot)   as a function of $g-r$ colour for different bins of stellar mass. The red dots are
the median age (metallicity) at fixed colour, while the error bars give the 68 percent percentile
range in the age (metallicity) distribution. Right panels: histogram of light-weighted age
(metallicity) compared to a Gaussian (dotted curve) of width equal to the average error on age
(metallicity) in each stellar mass bin. In each panel the rms scatter and the average error in
age (metallicity) are also given.}\label{CM_spread}
\end{figure*}

It is important to check that aperture effects do not introduce any spurious trend in
Fig.~\ref{CM_code}. This could arise because colour, luminosity and stellar mass  pertain to the
galaxy as a whole, while age, metallicity and $\rm \Delta(\mgtfe)$ are measured within a fixed
fibre aperture of radius 1.5 arcsec, which samples  different fractions of the total light in
galaxies with different apparent sizes. In section 3.4 of Paper~I, we already showed that
aperture effects are not expected to affect significantly the ages and metallicities derived for
the galaxies in our sample. To probe the influence of aperture effects on the results of
Fig.~\ref{CM_code}, we have arranged galaxies in several bins of apparent size $r_{50,r}$,
defined as the $r$-band Petrosian half-light radius. Galaxies with $1.5\la r_{50,r}\la2.4$~arcsec
are found out to redshift 0.2. We plotted the equivalent of Fig.~\ref{CM_code} for each bin of
$r_{50,r}$ within this range. In all cases, we found the same increase in metallicity, $\rm
\Delta(\mgtfe)$ and age along the relation as in Fig.~\ref{CM_code}, although the trends were
noisier  because of the smaller numbers of galaxies in individual $r_{50,r}$ bins. In our 
sample, galaxies along the CMR have a roughly constant apparent size $r_{50,r}\sim 2.3$~arcsec at
all magnitudes fainter than $M_r=-23.5$ (corresponding roughly to stellar masses less than
$10^{11.5} M_\odot$). The brightest galaxies tend to have larger  apparent sizes, but stellar
metallicity, age and $\rm \Delta(\mgtfe)$ do not show any significant trend with apparent size.
Thus, we are confident that aperture effects are not responsible for the trends seen in
Fig.~\ref{CM_code}.

For completeness, we repeated the analysis of Fig.~\ref{CM_code} using sample~B in  place of our
primary sample (Section~\ref{obscm}). We found no significant difference in this case with
respect to the results described above. We also repeated the analysis  using our primary sample,
but correcting the observed CMR for the effects of  dust attenuation (Section~\ref{obscm}).
Again, the main results in this case did not change. We further checked the influence of a
possible contamination of the trends identified in Fig.~\ref{CM_code} by late-type galaxies, in
particular, with regard to the variation in age along the relation. Excluding SF and C galaxies
(which tend to have younger ages  than the bulk of the sample) had no significant effect on the
results. In fact, C galaxies alone display the same trends as found in Fig.~\ref{CM_code} for the
sample as a whole.  The average metallicity of SF galaxies also increases with luminosity along
the CMR, although these tend to have light-weighted ages clustered around a mean value $\log(t_r/
{\rm yr}) =9.6$ almost independently of luminosity. We note that SF galaxies represent only about
5 percent of our primary sample, and they occupy $g-r$, $M_r$ bins where the fraction of galaxies
at a given magnitude is less than 2 percent of the total sample (not displayed in
Fig.~\ref{CM_code}). 

In summary, we have shown that the colour-magnitude relation is primarily a sequence in  galaxy
stellar mass. Both the chemical composition (i.e. the total metallicity and the
$\alpha$-elements-to-iron abundance ratio) and the age of elliptical galaxies depend mainly on
stellar mass, increasing along the relation. At the high-mass end of the relation, the age spread
is negligible and consistent with the errors. In this regime, the scatter in the CMR is
determined by the small scatter in the mass-metallicity relation. At lower masses, the
distribution in age becomes broader, with a spread toward younger ages, which correlates with
colour and is thus the main contributor to the scatter about the colour-magnitude relation at the
low luminosity end. 

\subsection{The \mgsig\ relation}\label{mgv}
In this section, we focus on another observational relationship between the stellar populations
and the structural properties of early-type galaxies: the relation between $\rm Mg_2$ index
strength and galaxy velocity dispersion $\rm \sigma_V$  \citep{bender93}. The difficulty in
interpreting this relation comes from the complex  translation of \mgtwo\ index strength and $\rm
\sigma_V$ into physical parameters. 

The \mgtwo\ index strengths considered here are corrected for broadening due to velocity
dispersion. They are  normalised to a fixed velocity dispersion of $\rm \sigv=200\,km~s^{-1}$,
corresponding  roughly to the average velocity dispersion of galaxies in our sample. The
normalization is achieved by using the \cite{bc03} models to compute the difference in \mgtwo\
index strength between the observed and reference velocity dispersion at the metallicity of each
galaxy.\footnote{This correction varies from $-0.002$ to $0.004$ over the range in velocity
dispersion covered by the sample. The average absolute correction is only 0.8 percent of the
$5-95$ percent percentile range in \mgtwo\ index strength.} For a consistent comparison with the
CMR and with the evolution-corrected light-weighted ages, we also correct the \mgtwo\ index
strengths so that they are relative to the present rather than to the redshift of observation, 
assuming passive evolution. The corrections are obtained by fitting simple linear relations
between \mgtwo\ index strength and time for SSPs of different metallicities and velocity
dispersions. The \mgtwo\ value measured for each galaxy is then corrected by adding the expected
change in index strength over a time interval equal to the look-back time at the redshift of the
galaxy.\footnote{The fitted slopes, averaged over velocity dispersion, are  $\rm
0.0056,0.0044,0.0064,0.0077~mag~Gyr^{-1}$ for $\log(Z/Z_\odot)=-0.7,-0.4,0,0.4$ respectively.}

Fig.~\ref{mg_V_dens} shows the \mgsig\ relation defined by the early-type  galaxies in our
primary sample. We have arranged galaxies in narrow bins of \mgtwo\ and  $\log\sigma_V$ (of
widths 0.01 and 0.036, respectively). By analogy with  Fig.~\ref{CM_dens}, the grey scale in
Fig.~\ref{mg_V_dens} indicates, for each $\log\sigma_V$ bin, the relative distribution of
galaxies in the different \mgtwo\ bins (we do not show bins containing less than 2 percent of the
total number of galaxies at given velocity dispersion). The solid straight line shows the result
of a robust linear fit obtained by minimising the absolute deviation in \mgtwo\ strength as a
function of $\log\sigma_V$.  This has a slope of $0.25\pm0.01$, i.e. very close to that found by
\cite{bernardi03d} for a sample of co-added, high-quality spectra of SDSS early-type galaxies
\citep[see also] []{guzman92,colless99}. The dashed lines in Fig.~\ref{mg_V_dens} are the mean
positive and negative absolute deviations in \mgtwo\ strength ($\pm0.024$ mag) relative to this
fit (also in agreement with previous work). These results are summarized in Table~\ref{fits}. We
also indicate in the table the slightly lower slope of the \mgsig\ relation obtained for sample~B
($0.19\pm0.002$). This is because sample~B includes fewer (emission-line)  galaxies with weak
\mgtwo\ absorption than our primary sample at low $\sigma_V$ and more  (low-S/N) galaxies with
weak \mgtwo\ absorption at large $\sigma_V$ (Section~\ref{obscm}).

\begin{figure}
\centerline{\includegraphics[width=8truecm]{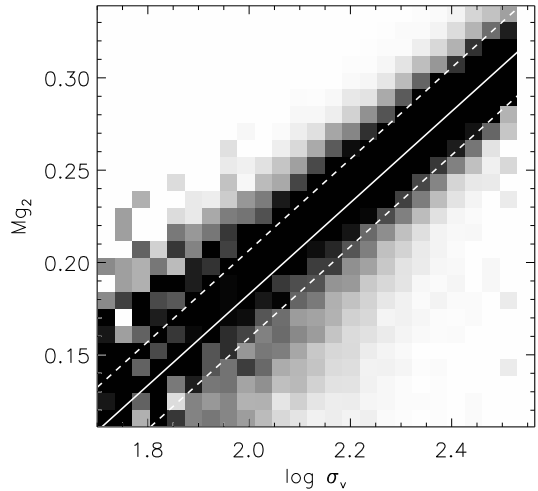}}
\caption{Relation between \mgtwo\ index strength (mag) and velocity dispersion $\log \sigma_V$
($\rm km~s^{-1}$) for the same sample of early-type galaxies as in Fig.~\ref{sample_distr}. As in
Fig. \ref{CM_dens}, the grey scale indicates, for each velocity dispersion bin, the relative
distribution of galaxies in the different $\rm Mg_2$ bins.}\label{mg_V_dens} \end{figure}

The tightness of the \mgsig\ relation has been used in the past to argue that early-type galaxies
have nearly coeval stellar populations, perhaps within $\sim$ 15  percent, at given composition
\citep[e.g.][]{bender93}. However, as noted by  \cite{trager00b}, the small scatter about the
relation could also conceal significant  age spreads, if these are accompanied by metallicity
spreads such that the $\rm Mg_2$  strength remains roughly constant at fixed velocity dispersion
\citep[see also][] {jorgensen99}. Assuming a moderate anticorrelation between age and
metallicity,  \cite{colless99} find that the intrinsic scatter in the \mgsig\ relation translates
into an upper limit of 40 (50) percent on the age (metallicity) spread. Moreover, it has been
pointed out that, at fixed $\sigma_V$, the distribution of \mgtwo\ residuals relative to the mean
\mgsig\ relation is not symmetric but skewed toward low $\rm Mg_2$ values, and that this effect
tends to increase at low velocity dispersion \citep{burstein88, bender93,jorgensen96,trager00b,
worthey03}. Our data confirm this finding. Fig. \ref{mgv_res} shows the distribution of \mgtwo\
residuals at different velocity dispersions (increasing from the top-left to the bottom-right
panels) for our primary sample. At  large velocity dispersion, the distribution of residuals is
symmetric and centred around  zero, while at smaller velocity dispersions, a tail of negative
residuals appears. This  could result from either an age or a metallicity spread, or both. The
observation that  galaxies with negative residuals are often morphologically disturbed
(presumably because of recent star formation) led to the idea that low $\rm Mg_2$ values at fixed
$\sigma_V$ are associated with younger ages \citep{ss92}. 

\begin{figure}
\centerline{\includegraphics[width=8truecm]{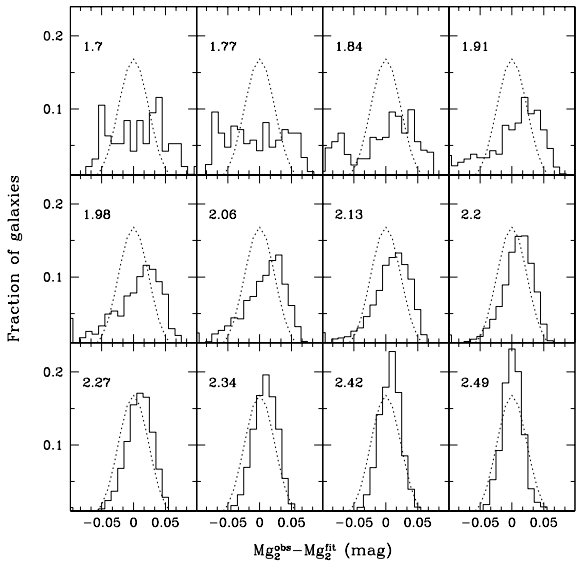}}
\caption{Distribution of residuals relative to the mean \mgsig\ relation fitted  in
Fig.~\ref{mg_V_dens}, for different values of $\log (\sigma_V/\rm km\,s^{-1})$ (indicated in each
panel).  The number of galaxies is normalised to the total in each bin of velocity  dispersion.
The dotted curve (repeated in each panel) is a Gaussian distribution of width equal to the
average absolute deviation in the \mgsig\ relation (see Table~\ref{fits}).}\label{mgv_res}
\end{figure}  

The availability of independent constraints on the ages and metallicities of early-type galaxies
in our sample allows us to re-examine the physical origin of the \mgsig\ relation.
Fig.~\ref{mgV_code} shows how metallicity, age, $\alpha$/Fe ratio and stellar mass change along
this relation. We have binned and colour-coded the $\rm Mg_2$, $\sigma_V$ plane to reflect the
average properties of the galaxies falling into each bin. As in  Fig.~\ref{mg_V_dens}, the
brightness of each $\rm Mg_2$ bin at a given $\sigma_V$ is  proportional to the number of
galaxies falling into this bin. Panels (a)--(c) show how (the median-likelihood estimates of)
age, metallicity and stellar mass are distributed along the \mgsig\ relation. Panel (d) shows the
distribution of the offset $\rm \Delta (\mgtfe)$ between observed and predicted \mgtfe\ index
strengths, which traces the  $\alpha$/Fe ratio (Section~\ref{sample}).

\begin{figure*}
\centerline{\includegraphics[width=12truecm]{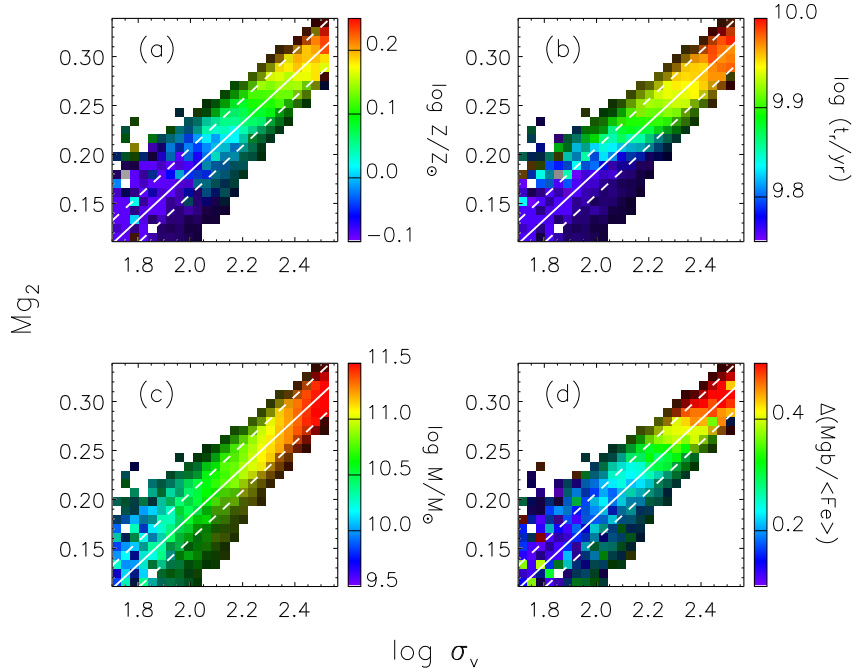}}
\caption{Relation between \mgtwo\ index strength (mag) and velocity dispersion ($\rm km~s^{-1}$)
for the same sample of early-type galaxies as in Fig.~\ref{sample_distr}. The relation has been
binned and colour-coded to reflect the average stellar metallicity (panel a), $r$-band
light-weighted age (panel b), stellar mass (panel c) and $\alpha$/Fe-estimator $\rm \Delta
(\mgtfe)$ (panel d) of the galaxies falling into each bin. The shading indicates the fraction of
galaxies populating each  \mgtwo\ bin a fixed  $\log \sigma_V$, according to the map of Fig.
\ref{mg_V_dens}.}
\label{mgV_code}
\end{figure*}

\begin{figure*}
\centerline{\includegraphics[width=8truecm]{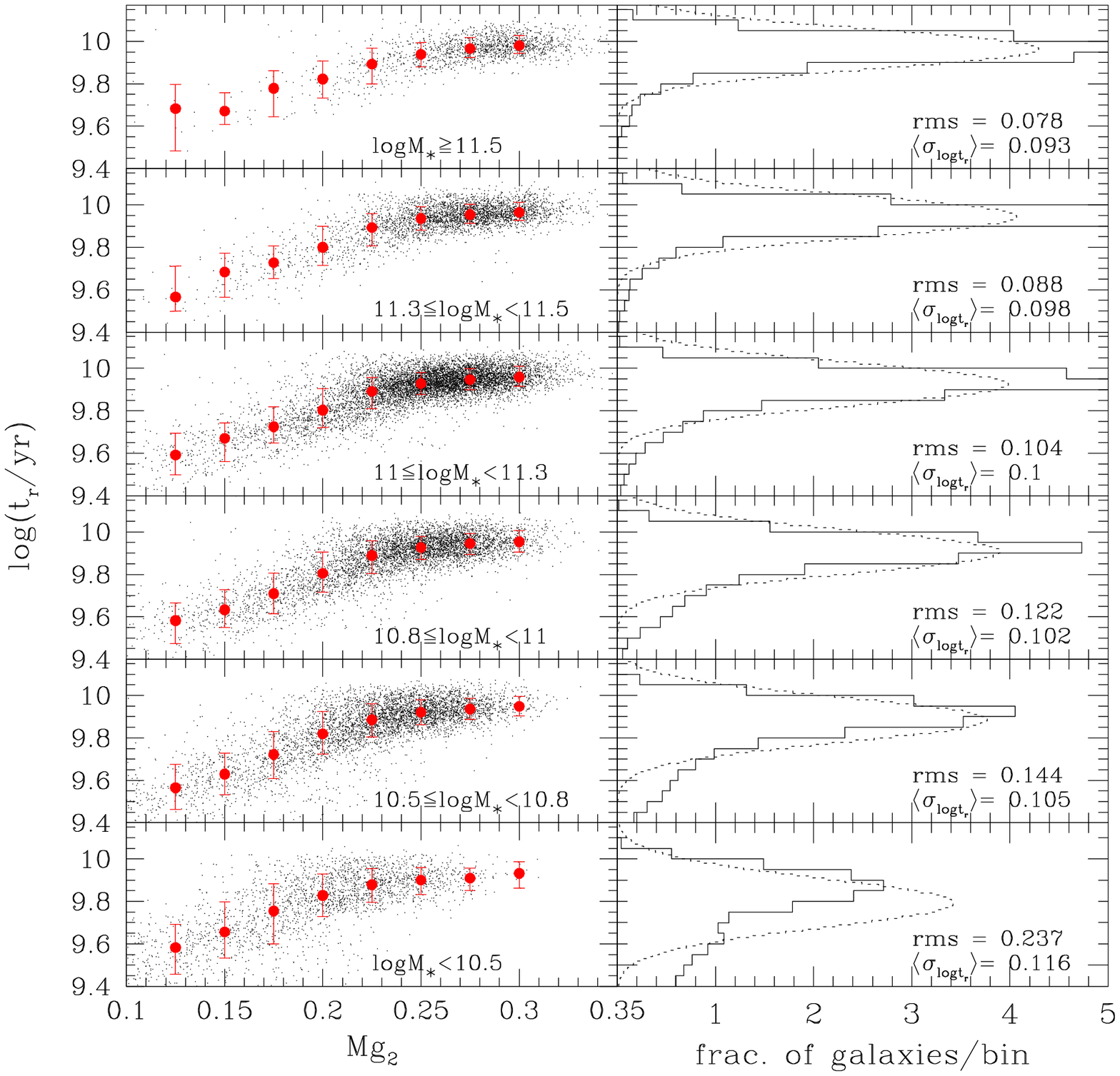}
\includegraphics[width=8truecm]{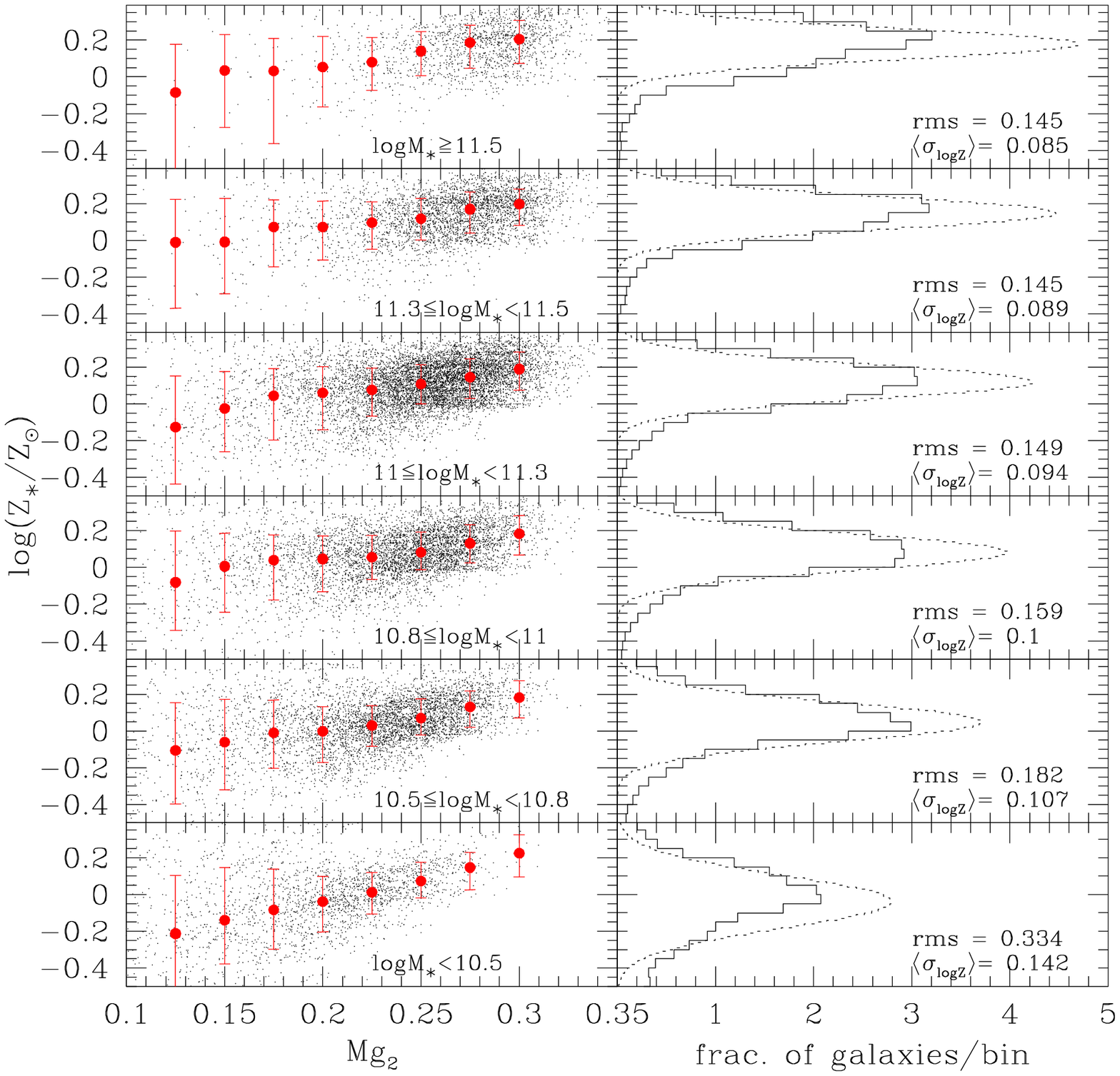}}
\caption{Left panels: light-weighted age (left plot) and stellar metallicity (right plot) as a
function of \mgtwo\ index strength (mag) in different bins of stellar mass. The red dots with
error bars represent the median and the 68 percent percentile range of the age (metallicity)
distribution at given \mgtwo\ index strength. Right panels: histogram of light-weighted age
(metallicity) compared to a Gaussian (dotted curve) of width given by the average error in age
(metallicity) for the galaxies falling into each stellar mass bin. The scatter and the average
error in age (metallicity) are also indicated.}\label{mgV_spread}
\end{figure*}

The most remarkable result from Fig.~\ref{mgV_code} is that, like the CMR, the \mgsig\ relation
for early-type galaxies appears to be primarily a sequence in stellar mass (panel c).  In fact,
stellar mass correlates tightly with velocity dispersion, which is a tracer of  dynamical mass
\citep[e.g.,][]{cappellari05}. Panel (a) further shows that stellar  metallicity increases along
the relation, from $\sim0.8 Z_\odot$ to $1.6Z_\odot$ from low to high velocity dispersions.
Galaxies with large $\rm \sigma_V$ are also older than  those with low $\rm \sigma_V$, though for
$\log\sigma_V \la 2.3$ age appears to correlate with \mgtwo\ index strength as well. Another
striking result from Fig.~\ref{mgV_code} is the similarity between the metallicity and $\rm
\Delta(\mgtfe)$ gradients along the \mgsig\ relation (panels a and d). This shows that massive
early-type galaxies with large velocity dispersions are both more metal-rich and more abundant in
$\alpha$ elements relative to iron than less  massive galaxies.

As in the case of the CMR (Section~\ref{physcm}), we have checked that the trends identified
above are not caused by aperture effects. In particular, we find similar trends in stellar mass,
metallicity, age and $\rm \Delta(\mgtfe)$ along the \mgsig\ relation when considering galaxies in
narrow ranges of $r_{50,r}$. Galaxies along the relation have roughly constant apparent size 
$r_{50,r}\sim 2.3$~arcsec, implying that the average fraction of total galaxy flux sampled by the
fibre is almost constant.

As above for the CMR, we can quantify the relative contribution of metallicity and age to the
scatter in the \mgsig\ relation. We note that there must be another parameter, most likely the
$\alpha$-elements-to-iron abundance ratio, that is responsible for the scatter in the relation.
In fact, the scatter in \mgtwo\ of [$\alpha$/Fe]=0 models with the ages and metallicities of the
galaxies in our sample is only $\pm0.0118$, i.e. much smaller than the observed scatter
(0.0237).\footnote{For each galaxy in the sample, we adopt here the median of the likelihood
distribution in \mgtwo\ obtained by weighting each model by $\chi^2=[(\log Z_{mod} -\log
Z)/\sigma_{\log Z}]^2+[(\log t_{mod}-\log t_r)/\sigma_{\log t_r}]^2$.} We obtain a similar
scatter ($\pm0.0116$) in the \mgtwo\ index strength of models with the same metallicity as the
galaxies in our sample but with a fixed imposed (average) age at fixed $\log\sigma_V$. On the
other hand the scatter in \mgtwo\ index strength of models with the ages of the galaxies in our
sample but with a fixed imposed (average) metallicity at fixed $\log\sigma_V$ is only $\pm0.007$.
This indicates that metallicity has a stronger influence than age on the scatter in the \mgsig\
relation. 

Figure~\ref{mgV_spread} is obtained in a similar way as Fig.~\ref{CM_spread}. As in
Fig.~\ref{CM_spread}, the left-hand panels of Fig.~\ref{mgV_spread} show that the scatter in
light-weighted age of high-mass ellipticals ($M_\ast\ge10^{11}M_\odot$) is negligible and
consistent with the errors on the age estimates. Only at masses below $10^{11}M_\odot$ does a
significant tail of younger ages appear. These young galaxies are responsible for the scatter in
\mgtwo\ at fixed mass, but the relation between age and index strength saturates for
$\mgtwo\ga0.25$. Also in agreement with Fig.~\ref{CM_spread}, the right-hand plot shows that we
detect (above the measurement error) a small scatter in metallicity at fixed mass in all mass
bins. There is a correlation between metallicity and \mgtwo\ at fixed mass, in particular for
galaxies with $\mgtwo\ga0.25$ (left-hand panels), where light-weighted age saturates to a
constant mean value.

The \mgsig\  relation is, therefore, primarily a sequence in galaxy stellar  mass. It reflects
the fact that early-type galaxies form a sequence of increasing total  stellar metallicity and
$\alpha$-elements to iron abundance ratio from shallow to  deep potential wells. Our results also
confirm the trend of increasing age with increasing velocity dispersion
\citep[e.g.][]{caldwell03, thomas05,nelan05}. At high masses, the small scatter in the relation
correlates with the scatter in stellar metallicity at fixed stellar mass, while light-weighted
age is almost independent of index strength. At the low-mass end of the relation, the fraction of
young ellipticals increases and variations in age between $\sim4$ and $\sim8$ Gyr correlate with
index strength at fixed mass.

\begin{table}
\caption{Parameters of the relations fitted by minimising the absolute vertical
deviations. Sample A refers to our primary sample, while sample B is obtained by
selecting galaxies according to \citet{bernardi05} criteria.}\label{fits}
\centering
\begin{tabular}{l|ccc}
\hline
Sample & Slope & Intercept & Scatter \\
\hline
\multicolumn{4}{c}{Colour-magnitude relation}\\
\hline
sample A$^{\it a}$ & $-0.015\pm0.001$ & 0.40 & 0.051 \\
sample B$^{\it a}$ & $-0.014\pm0.002$ & 0.41 & 0.043 \\
sample A$^{\it b}$& $-0.024\pm0.002$ & 0.22 & 0.050 \\
sample B$^{\it b}$& $-0.028\pm0.001$ & 0.14 & 0.043 \\
sample A$^{\it c}$& $-0.025\pm0.001$ & 0.14 & 0.051 \\
sample B$^{\it c}$& $-0.015\pm0.002$ & 0.37 & 0.043 \\
sample A$^{\it d}$& $-0.035\pm0.002$ & $-0.05$ & 0.050 \\
sample B$^{\it d}$& $-0.029\pm0.002$ & 0.09 & 0.044 \\
\hline
\multicolumn{4}{c}{\mgsig\ relation}\\
\hline
sample A$^{\it e}$ & $0.23\pm0.01$ & $-0.29$ & 0.024\\
sample B$^{\it e}$ & $0.17\pm0.003$ & $-0.15$ & 0.023 \\
sample A$^{\it f}$ & $0.25\pm0.01$ & $-0.31$ & 0.024\\
sample B$^{\it f}$ & $0.19\pm0.00$ & $-0.18$ & 0.023 \\
\hline	  
\end{tabular}
\begin{minipage}{8.0truecm}
\footnotesize{$^{\it a}${\it k}-corrected colour and magnitude.}
\footnotesize{$^{\it b}${\it k}+{\it e}-corrected colour and magnitude \citep[the evolution correction is the
one provided by][]{bernardi05}.}
\footnotesize{$^{\it c}${\it k}-corrected colour and magnitude dereddened applying the
average dust corrections of Fig.~\ref{CM_dust}.}
\footnotesize{$^{\it d}$Fully corrected colour and magnitude.}
\footnotesize{$^{\it e}$\mgtwo\ index strength corrected for velocity dispersion.}
\footnotesize{$^{\it f}$\mgtwo\ index strength corrected for velocity dispersion and evolution.}
\end{minipage}
\end{table}

\subsection{Environmental dependence}\label{environment}
Several authors have addressed the dependence on environment of the properties of the stellar
populations in early-type galaxies. Studies of the colour-magnitude relation and the relations
between absorption indices and velocity dispersion on relatively small samples of early-type
galaxies in different environments have shown that galaxies in low-density environments tend to
be younger and more metal-rich than those in high-density environments
\citep[e.g.][]{kunt02,thomas05,denicolo05}. \cite{bernardi98}, later confirmed by
\cite{bernardi06} on a larger sample of SDSS early-type galaxies, found differences in the
\mgtwo$-$\sigv\ relation of galaxies in different environments, implying that galaxies in dense
environments are at most 1 Gyr older than galaxies in low-density environments and that they have
the same metallicity. 

\cite{kauf04} provide estimates of environmental density for a sample of SDSS DR2 galaxies in the
redshift range $0.03<z<0.1$ and with apparent $r$-band magnitude in the range $14.5<r<17.77$,
complete down to a stellar mass of ~$2\times10^9\rm M_\odot$. The density is expressed in terms
of the number of spectroscopically-observed neighbouring galaxies (down to a fixed absolute
magnitude) within 2~Mpc of projected radius and $\pm500$~km/s in velocity difference from the
target galaxy, corrected for galaxies missed due to fibre collisions ($\rm N_{neigh}$). We take
advantage of these density estimates to address any environmental dependence of the physical
properties of the stellar populations for the galaxy sample studied here. This can be achieved
for only 1765 galaxies in our sample, for which an estimate of $\rm N_{neigh}$ is available. We
consider three bins in environmental density, defined by $\rm N_{neigh}<4$, $\rm 4\leq
N_{neigh}<7$ and $\rm N_{neigh}\geq 7$, which contain, respectively, 693, 388 and 684 galaxies.
As mentioned in Section~\ref{sample}, we can classify the galaxies in our sample on the basis of
their emission-line properties. As expected, the sample is dominated by `unclassifiable' galaxies
(without emission lines), but there is also contamination by galaxies with a low level of star
formation (SF and C galaxies). Fig.~\ref{class_env} illustrates the fraction of unclassifiable,
SF, low-S/N SF, C and AGN galaxies as a function of environment. This plot quantifies the
statement of Section~\ref{obscm} that the fraction of galaxies showing emission lines in our
sample increases in lower-density environments. This class of galaxies also contributes to
increase the scatter blueward of the colour-magnitude relation.

In Fig.~\ref{rel_env}, we explore how the CMR (left-hand panels) and the \mgtwo$-$\sigv\ relation
(right-hand panels) depend on environment. The relations found in the highest-density bin are
compared to those defined in the lowest-density bin. The results of the linear fits (also for the
intermediate bin of $\rm N_{neigh}$) are given in Table~\ref{fits_env}. Fig.~\ref{rel_env} and
Table~\ref{fits_env} show that there is no systematic variation in the slope of the CMR as a
function of environment, while the \mgtwo$-$\sigv\ relation appears to steepen at low densities
because of a larger fraction of galaxies with low \mgtwo\ index strength  at low velocity
dispersions (columns 3 and 6). Between the two extreme density bins there are differences of
$0.006\pm0.003$ and $0.007\pm0.003$ in the zero-points of the CMR and the \mgtwo$-$\sigv\
relation,  respectively (columns 4 and 7). This is in agreement with the small shift of
$0.007\pm0.002$ mag measured by \cite{bernardi98} in the \mgtwo$-$\sigv\ relation of a sample of
931 early-type galaxies in different environments. We also identify a systematic increase of the
scatter about both relations from high- to low-density environments, in agreement with earlier
findings, as mentioned above \citep[e.g.][]{hogg04}.

It is of interest to understand how the change in the scatter along the two scaling relations
reflects differences in the physical parameters of galaxies in different environments. We note
that the distribution in stellar mass does not vary significantly\footnote{When comparing the
stellar mass distribution of galaxies in the low-density bin with that of galaxies in the
high-density bin, the probability obtained from a Kolmogorov-Smirnov test is not low enough to
reject the hypothesis that the two distributions are drawn from the same parent distribution.}
with environment, but that the median stellar mass of galaxies in the lowest-density bin is lower
by about 0.05~dex than the median stellar mass in the highest-density bin (it increases from
$8\times10^{10}$ to $9\times10^{10}\rm M_\odot$). Since stellar metallicity, age and element
abundance ratio all increase with stellar mass, any effect induced by changes in stellar mass
must be removed when quantifying  variations in these parameters with environment.  To do this,
we calculate the median stellar metallicity, light-weighted age and element abundance ratio as a
function of stellar mass for the sample as a whole (see Fig.~\ref{par_sigma}). For the 1765
galaxies with an estimate of environmental density, we then consider the offsets in
$\log(Z/Z_\odot)$, $\log(t_r/yr)$, $\Delta(\mgtfe)$ from the median values of these parameters at
fixed stellar mass in the whole sample. The distributions in $\Delta[\log Z]$ and $\Delta[\log
t_r]$ are skewed toward negative values, independently of environment, and this effect is
stronger at small masses. If galaxies at low densities are distributed preferentially to smaller
masses than galaxies in dense environments, the distributions in $\Delta[\log Z]$ and
$\Delta[\log t_r]$ will show a stronger tail toward negative values in the low-density bin. To
separate this effect from an intrinsic dependence of metallicity and age on environment, we
further distinguish between galaxies with stellar masses above and below $\rm 10^{11}M_\odot$.

The result of this analysis is shown in Fig.~\ref{distr_env} for the same three environmental
bins as considered above. The distributions of the offsets in $\log(Z/Z_\odot)$, $\log(t_r/yr)$
and $\rm \Delta(\mgtfe)$ for the two low-density bins ($\rm N_{neigh}<4$ and $\rm 4\leq
N_{neigh}<7$, solid lines) are compared to the corresponding ones in the high-density bin ($\rm
N_{neigh}\geq 7$, dotted line in each panel), for massive and low-mass galaxies separately (red
and blue histograms, respectively). The comparison is also summarized in Table~\ref{stat_env},
where we give the difference of the average parameter offset between the highest- and the
lowest-density bins, for galaxies with stellar mass above and below $\rm 10^{11}M_\odot$
separately. There, we also indicate the probability for the two distributions to be drawn from
the same parent distribution, according to a Kolmogorov-Smirnov test. Element abundance ratio, as
expressed by $\rm \Delta(\mgtfe)$, does not show any significant variation with environment,
either in the average value or in the scatter. In contrast, light-weighted age (independently of
mass) and metallicity (for massive galaxies) show a small dependence on environment in the sense
that there is a higher fraction of young, metal-poor galaxies in low-density environments.
Massive early-type galaxies in dense environments tend to be $\sim0.03$~dex more metal-rich than
their field counterparts. Similarly, the light-weighted ages of galaxies in dense environments
are $\sim0.02$~dex older than in the field. It is interesting to mention that we also find a
systematic increase in the scatter of metallicity and age from dense to low-density environments.
From the highest- to the lowest-density bins the scatter in both metallicity and age increases by
about 0.02~dex for massive galaxies (from 0.118 to 0.135 for $\log (Z/Z_\odot)$, from 0.083 to
0.103 for $\log (t_r/yr)$) and by about 0.01~dex for $M_\ast<10^{11}\rm M_\odot$ (from 0.16 to
0.17 for $\log (Z/Z_\odot)$ and from 0.12 to 0.136 for $\log (t_r/yr)$). Although very small,
these trends hint at a possibly very relevant environmental dependence of metallicity and age.
Future analysis of larger samples of galaxies (provided, e.g., by the complete SDSS) with well
characterized environmental properties will allow to draw firmer conclusions. 

\begin{figure}
\centerline{\includegraphics[width=6truecm]{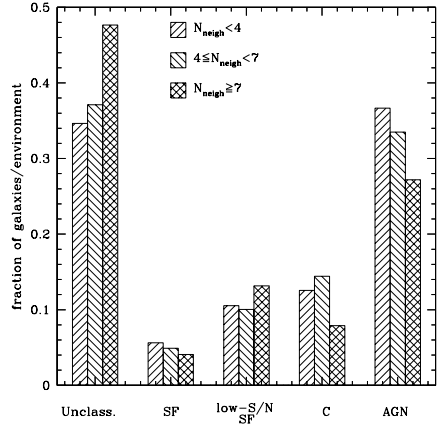}}
\caption{Fraction of unclassifiable ('Unclass'), star-forming ('SF'), low-S/N star-forming
('low-S/N SF'), composite ('C') and AGN galaxies as a function of environment. Three different
bins of environmental density are considered here, as given by $\rm N_{neigh}$ in the plot. Each
histogram is normalised to the number of galaxies in the corresponding environmental bin. While
the high-density environments are strongly dominated by galaxies without emission lines
(unclassifiable), the fraction of star-forming, composite and AGN galaxies increases in
lower-density environments.}\label{class_env}
\end{figure}

\begin{figure*}
\centerline{\includegraphics[width=8truecm]{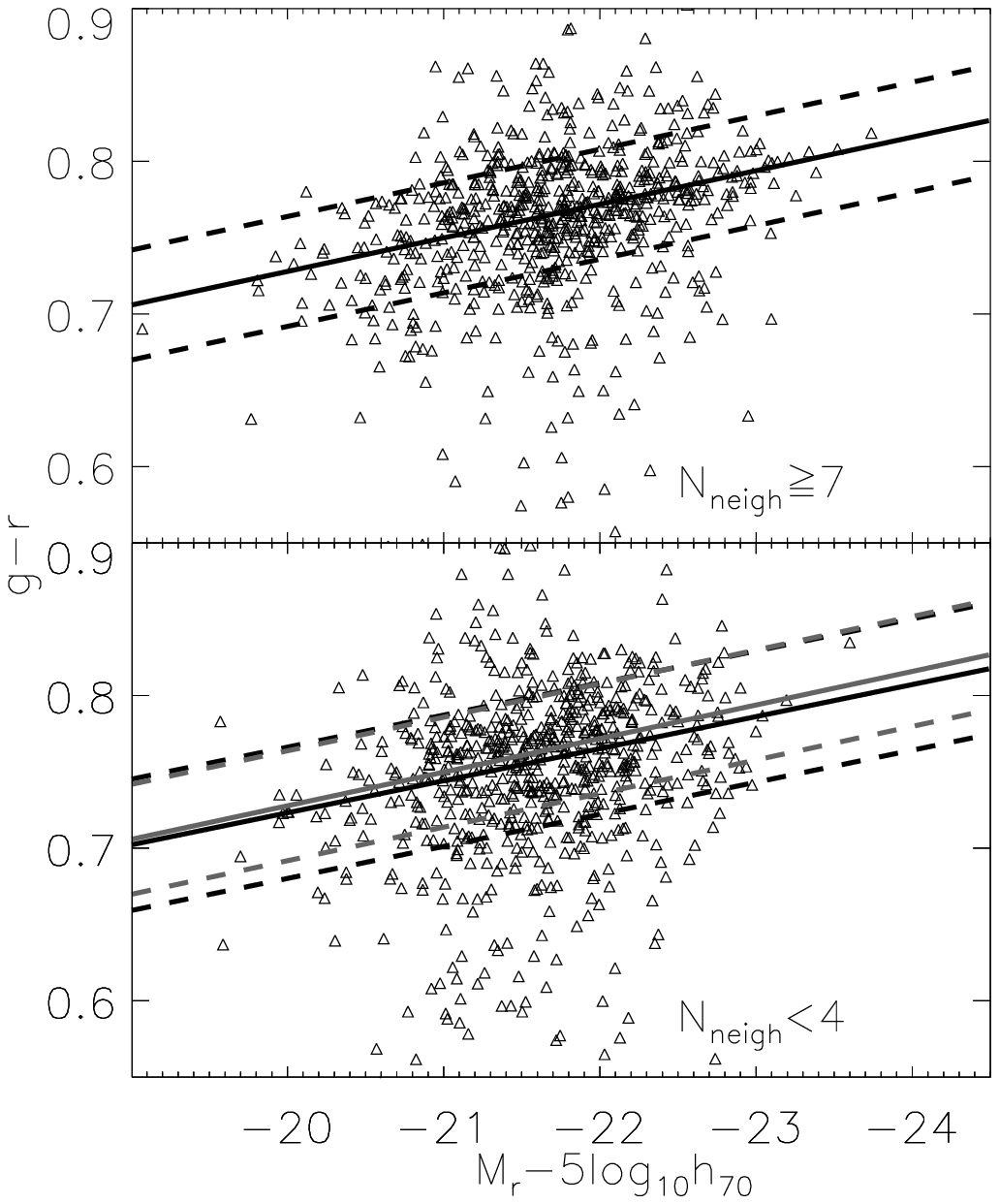}
\includegraphics[width=8truecm]{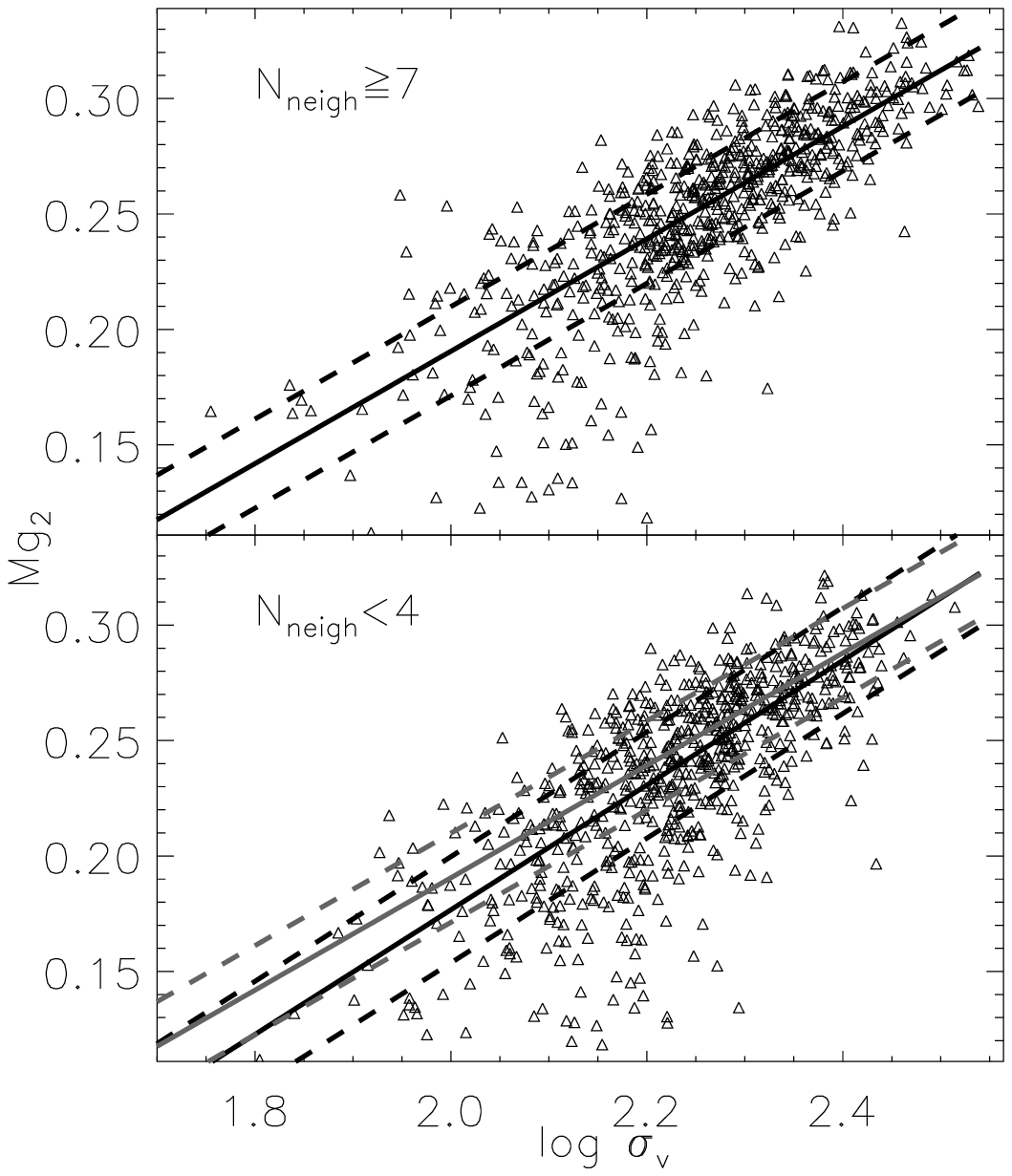}}
\caption{Colour-magnitude (left plot) and \mgtwo$-$\sigv\ relations (right plot) as a function of
environment. Galaxies are divided into two disjoint bins of environmental density (increasing
from bottom to top as indicated by $\rm N_{neigh}$ in each panel). The linear relation fitted for
each subsample and the scatter about it are shown in each panel by the black solid and dashed
lines. The grey lines reproduce the relation fitted in the highest density bin.}\label{rel_env}
\end{figure*}

\begin{table*}
\caption{Parameters of the colour-magnitude relation (columns 3,4,5) and \mgtwo$-$\sigv\ relation
(columns 6,7,8) for a subsample of 1765 galaxies with environmental estimates. The relations are
fitted separately for three different bins of environmental density (given by $\rm N_{neigh}$,
column 1). The number of galaxies in each bin is given in column 2. The intercepts are given at
$M_r=-21.5$ and $\log\sigv=2.25$, which correspond roughly to the average magnitude and velocity
dispersion of this subsample of galaxies.}\label{fits_env}
\centering
\begin{tabular}{lc|ccc|ccc}
\hline
 & &\multicolumn{3}{c}{colour-magnitude} & \multicolumn{3}{c}{$\rm Mg_2-\sigma_V$} \\
\hline
Environment & $\rm N_{gal}$ & Slope & Intercept & Scatter & Slope & Intercept & Scatter \\
\hline
\multicolumn{1}{c}{(1)} & (2) & (3) & (4) & (5) & (6) & (7) & (8) \\
\hline
$\rm N_{neigh}<4$       & 693 & $-0.021\pm0.004$ & $0.755\pm0.002$ & 0.043 & $0.27\pm0.03$ & $0.244\pm0.002$ & 0.023 \\ 
$\rm 4\leq N_{neigh}<7$ & 388 & $-0.026\pm0.006$ & $0.757\pm0.004$ & 0.041 & $0.26\pm0.01$ & $0.242\pm0.001$ & 0.020 \\
$\rm N_{neigh}\geq 7$	& 684 & $-0.022\pm0.003$ & $0.761\pm0.003$ & 0.036 & $0.24\pm0.02$ & $0.251\pm0.003$ & 0.019 \\
\hline
\end{tabular}
\end{table*}

\begin{figure}
\centerline{\includegraphics[width=8truecm]{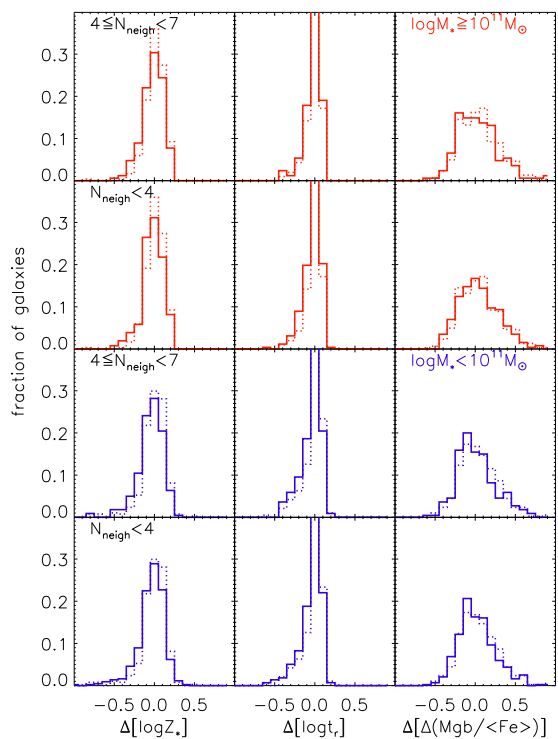}}
\caption{Stellar metallicity, age and element abundance ratio (expressed as $\rm
\Delta(\mgtfe)$)  as a function of environment for the subsample of 1765 galaxies with
environmental estimates. We have estimated the offset in $\log(Z/Z_\odot)$, $\log(t_r/yr)$ and
$\rm \Delta(\mgtfe)$ from the median value at fixed stellar mass calculated for the sample as a
whole. The distributions in the offsets are more skewed toward negative values at smaller masses.
Thus, to further remove any effect induced by an environmental dependence of the mass, we
analysed separately galaxies with masses above and below $\rm 10^{11}M_\odot$ (shown in red and
blue respectively). For each stellar mass bin, the distributions in the offsets are shown for
galaxies with different environmental density ($\rm N_{neigh}<4$ in the lower panels and $\rm
4\leq N_{neigh}<7$ in the upper panels, solid line). Each distribution is compared to that for
galaxies with $\rm N_{neigh}\geq7$, shown by the dotted line in each panel.}\label{distr_env}
\end{figure}

\begin{table}
\caption{Difference between the average parameter offset between the $\rm N_{neigh}<4$ and $\rm
N_{neigh}\geq7$ environmental bins, distinguishing between galaxies with stellar masses above and
below $\rm 10^{11}M_\odot$. For each parameter, the second row gives the probability from a
Kolmogorov-Smirnov test that the two distributions are consistent with being drawn from the same
parent distribution (we consider differences to be statistically significant only if the KS
probability is below 1 percent).}\label{stat_env}
\centering
\begin{tabular}{l|cc|}
\hline
 & $M_\ast\geq10^{11}\rm M_\odot$ & $M_\ast<10^{11}\rm M_\odot$ \\
\hline
$\Delta[\log Z]$ & $0.032\pm0.006$ & $0.029\pm0.006$ \\
                 & 0.0030          & 0.0217          \\
\hline
$\Delta[\log t_r]$ & $0.019\pm0.005$ & $0.023\pm0.006$ \\
                   & 0.0053          & 0.0017          \\
\hline
$\Delta[\Delta(\mgtfe)]$ & $0.008\pm0.009$ & $0.024\pm0.009$ \\
                         & 0.5321          & 0.1499          \\
\hline 
\end{tabular}
\end{table}

\subsection{Correlations between physical parameters}\label{mass}
So far, we have considered separately the constraints set by the colour-magnitude and the
$\mgsig$ relations on the ages, chemical compositions and stellar masses of early-type galaxies
in our sample. Here, we explore in more detail  the potential correlations between these
different physical parameters, the dynamical mass and the stellar surface mass density of
galaxies.

We first turn our attention to the relation between age, luminosity, velocity dispersion and 
stellar mass inferred from the distribution in the full parameter space described by the
colour-magnitude and the $\mgsig$  relations. Figs~\ref{CM_code}b and \ref{mgV_code}b taken
together indicate that brighter galaxies or galaxies with large velocity dispersions are on
average older than fainter or smaller velocity dispersion ones (consistent with the `downsizing'
scenario; \citealt{cowie96}, see also  Section~\ref{summary}). In Fig.~\ref{lum_V}a, we show  the
relation between $M_r$ absolute magnitude, velocity dispersion $\sigma_V$ and  light-weighted age
for the galaxies in our primary sample. At fixed luminosity, galaxies with large $\sigma_V$ tend
to be older than those with small $\sigma_V$ (a result already pointed out by \citealt{FP99} and
\citealt{bernardi05}). Conversely, at fixed velocity dispersion, luminous galaxies tend to be
younger than faint galaxies. The dispersion in the $M_r$-$\sigma_V$  relation and the dependence
of luminosity on age at fixed $\sigma_V$ cannot be accounted for  entirely by the fact that the
stellar mass-to-light ratio $M_\ast/L_r$ of an evolving stellar  population increases with
light-weighted age. This is shown in Fig.~\ref{lum_V}b, where we  plot the relation between
stellar mass and velocity dispersion for the same galaxies as in  Fig.~\ref{lum_V}a. The scatter
of 0.252 dex about this relation is consistent with the scatter  of 0.579~mag in the relation
between $M_r$ and $\sigma_V$. There appears to be a real dependence  of age on stellar mass at
fixed velocity dispersion, in the sense that galaxies with large  stellar masses tend to be
slightly younger than those with low stellar masses (Table~\ref{fits_par}  provides simple linear
fits to the relations shown in Figs~\ref{lum_V}a and b).

\begin{figure}
\centerline{\includegraphics[width=8truecm]{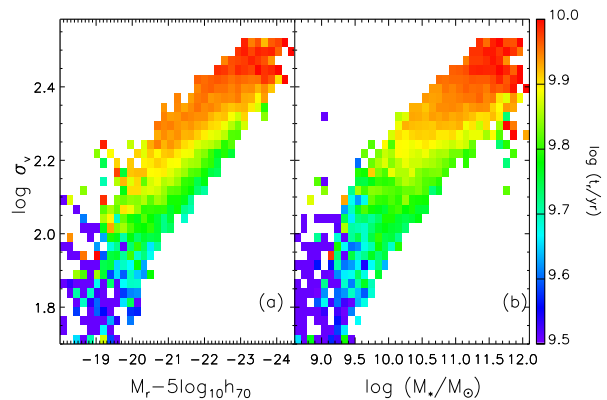}}
\caption{Relation between velocity dispersion and $r$-band absolute magnitude (panel a),
colour-coded to reflect the average light-weighted age of the galaxies falling into each $\log
\sigma_V$--$M_r$ bin. For $M_r$ brighter than $-20$, lines of constant age are approximately
parallel to the relation. Panel (b) shows the result of substituting absolute magnitude with
stellar mass.}\label{lum_V}
\end{figure}

\begin{table*} 
\caption{Linear fits to the relations plotted in Figs~\ref{lum_V} and \ref{par_sigma}.}\label{fits_par} 
\centering 
\begin{tabular}{l|ccc} 
\hline 
\hline
& Slope &Intercept & Scatter \\ 
\hline 
\hline
$\log\sigv$ vs $M_r$ & $-0.125\pm0.004$ & $-0.464$ & 0.073 \\
$\log\sigv$ vs $\log M_\ast$ & $0.286\pm0.020$ & $-0.895$ & 0.071 \\ 
\hline
$\log t_r$ vs $\log M_{dyn}$ & $0.115\pm0.056$ & 8.628 & 0.085 \\ 
$\log t_r$ vs $\log M_\ast$  & $0.112\pm0.098$ & 8.678 & 0.088 \\
\hline 
$\log Z$ vs $\log M_{dyn}$ & $0.164\pm0.010$ & $-1.731$ & 0.119 \\ 
$\log Z$ vs $\log M_\ast$ & $0.168\pm0.013$ & $-1.757$ & 0.122 \\ 
\hline
$\Delta(\mgtfe)$ vs $\log M_{dyn}$ & $0.165\pm0.012$ & $-1.517$ & 0.199 \\ 
$\Delta(\mgtfe)$ vs $\log M_\ast$ & $0.128\pm0.013$ & $-1.107$ & 0.204 \\ 
\hline
\hline	   
\end{tabular} 
\end{table*}

The results of Fig.~\ref{lum_V} motivate us to investigate more closely the relation between
stellar mass and dynamical mass for early-type galaxies. We can estimate the dynamical mass
(including stars and dark matter) within the optical radius of a galaxy from a combination of
velocity dispersion and optical size \citep{cappellari05}. Here we use the radius  containing 50
percent of the Petrosian flux in the $r$-band (\rpetro) rather than the  effective de Vaucouleurs
radius ($r_{deV}$), because we do not explicitly select early-type galaxies on the basis of the
shape of their light profile. Thus, a de Vaucouleurs profile  may not always be an optimal fit to
the photometric data, and the empirical half-light  radius \rpetro\ is more straightforward to
interpret.\footnote{The de Vaucouleurs effective radii of the galaxies in our sample are on
average $\sim1.4$ times larger than the Petrosian half-light radii. This arises presumably from
the fact that the Petrosian flux represents  typically 80 percent of the total flux in the de
Vaucouleurs model \citep{padman04}, but this  could also result (at least for a class of objects)
from forcing a de Vaucouleurs fit to a flatter galaxy light profile.} To estimate the dynamical
mass within \rpetro, we must also account for  the fact that the velocity dispersion of SDSS
galaxies is measured within a fixed 1.5~arcsec  fibre radius. Following \cite{jorgensen99}, we
assume that the radial profile of the velocity dispersion has a slope $-0.04$. The velocity
dispersion at the radius \rpetro\ is then given by
\begin{equation}
\frac{\sigma_V(r_{fibre})}{\sigma_V(\rpetro)} = \left(\frac{r_{fibre}}{\rpetro}\right)^{-0.04},
\end{equation}
and the dynamical mass within \rpetro\ by
\begin{equation}
\frac{M_{dyn}}{M_\odot} = \xi \frac{\sigma_V^2(\rpetro)~\rpetro}{G}.
\label{mdyn}
\end{equation}
In this expression, $G$ is the gravitational constant, and $\xi$ is a scaling factor given in
Table~\ref{mass_fits}, which depends on the shape of the velocity dispersion  profile (this
factor would be equal to 5 if we estimated the dynamical mass within the effective de Vaucouleurs
radius; see \citealt{cappellari05}).

\begin{figure}
\centerline{\includegraphics[width=8truecm]{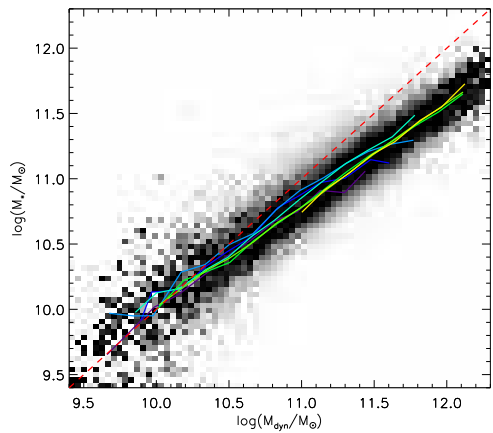}}
\caption{Relation between stellar mass and dynamical mass M$_{dyn}$ estimated within the
$r$-band  Petrosian half-light radius (see equation~\ref{mdyn}). Lines of different colours
represent the  median relations in different bins of light-weighted age, increasing from
$\log(t_r/{\rm yr})=9.5$ (purple) to 10.1 (yellow). The ratio between stellar and dynamical mass
decreases with mass, as highlighted by the comparison with the one-to-one relation (dashed
line).}\label{masses}
\end{figure}

In Fig.~\ref{masses}, we compare the dynamical masses obtained in this way to the stellar masses
derived in Paper~I for the early-type galaxies in our primary sample.\footnote{We recall that 
the total stellar mass is obtained by multiplying the stellar mass-to-light ratio derived from a
fit of the fibre spectrum by the total luminosity of the galaxy. This assumes that the stellar
mass-to-light ratio outside the fibre is the same as that inside the fibre (see Paper I).} We
focus on the slope of the relation, because the zero point is sensitive to our assumptions about
the radius within which the dynamical mass is estimated, the conversion factor between dynamical
and virial masses, and the  stellar initial mass function (IMF; which we assume constant for all
the galaxies). A linear fit  to the relation in  Fig.~\ref{masses} yields
\begin{equation}
(M_\ast/M_\odot)\propto(M_{dyn}/M_\odot)^{(0.783\pm0.019)}
\label{masseq}
\end{equation}\label{dynmass}
with a scatter of $0.13$~dex. The specific correction applied to $\sigma_V$ has little effect on
the fitted slope. As a check, we tried steeper velocity dispersion profiles, with slopes $-0.06$
\citep{mehlert03} and $-0.066$ \citep{cappellari05}. These yielded relations between $\log
M_\ast$ and $\log M_{dyn}$ with slopes $0.785\pm0.021$ and $0.786\pm0.017$, respectively, i.e.
consistent within 1$\sigma$ with the slope of the relation in Fig.~\ref{masses}. We also tested
the effect of estimating the dynamical mass within the effective de Vaucouleurs radius $r_{deV}$
instead of \rpetro. The dynamical masses obtained in this case were systematically higher, by
$\sim$0.13~dex,  than those derived within \rpetro. The relation between $\log M_\ast$ and $\log
M_{dyn}$ had a slope of  $0.808\pm0.026$, i.e. slightly higher but consistent within 1$\sigma$
with the slope of the relation in Fig.~\ref{masses}. These results are summarised in
Table~\ref{mass_fits}. 

\begin{table}
\caption{Correlation between stellar and dynamical mass. The first two columns give the radius
within which the dynamical mass is estimated and the slope of the velocity dispersion profile 
assumed to correct for aperture effects.}\label{mass_fits}
\centering
\begin{tabular}{ccc|ccc}
\hline
Radius & $\sigma_V$ profile & $\xi$ & Slope & Intercept & Scatter \\
\hline
\rpetro\ & $-0.040$ & 7.14 & $0.783\pm0.019$ & 2.19 & 0.127 \\
\rpetro\ & $-0.060$ & 6.98 & $0.785\pm0.021$ & 2.18 & 0.128 \\
\rpetro\ & $-0.066$ & 7.02 & $0.786\pm0.017$ & 2.17 & 0.128 \\
$r_{deV}$ & $-0.040$ & 5.00 & $0.808\pm0.026$ & 1.93 & 0.113 \\
\hline          
\end{tabular}
\end{table}

A robust result from Fig.~\ref{masses}, therefore, is that the ratio between dynamical mass and
stellar mass increases from the least massive to the most massive early-type galaxies in our
sample. Structural non-homology does not appear to be responsible for this effect. We have built
different subsamples of galaxies, based on the value of the Sersic index $n$ fitted to the light
profile in the SDSS database. The slope of the relation between $\log M_\ast$ and $\log M_{dyn}$
within each subsample remains close to that of the relation in Fig.~\ref{masses}, changing from
0.847 for $n=3$ to 0.801 for $n=5.5$ (a de Vaucouleurs profile corresponding to $n=4$). Instead,
the decrease in $M_\ast/ M_{dyn}$ with stellar mass in Fig.~\ref{masses} is consistent with the
increase in the dynamical  mass-to-light ratio ($M_{dyn}/L$) of early-type galaxies implied by
the Fundamental Plane under the assumption of structural homology
(\citealt{bender92,pierini02,zibetti02}; see also  \citealt{cappellari05}, where no assumption on
homology is made). This is also consistent with the increase in $M_{dyn}/L$ with luminosity 
found by \cite{padman04} for a sample of 29,469 SDSS elliptical galaxies. The decrease of
$M_\ast/ M_{dyn}$ with stellar mass could result from a more efficient mixing of dark matter and
stars within the optical radius of massive galaxies relative to low-mass galaxies, as expected if
the most massive early-type galaxies assembled through multiple mergers of dissipationless
systems \citep[see for  discussion][]{white80,humphrey05,delucia05,boylan-kolchin05}.

The trend in $M_\ast/M_{dyn}$ with stellar mass shows a weak dependence on galaxy light-weighted
age. This is shown in Fig.~\ref{masses}, where lines of different colours indicate the median
stellar mass as a function of dynamical mass for galaxies in various age bins, from
$\log(t_r/{\rm yr})= 9.5$ (purple) to 10.1 (yellow). Lines of constant age run parallel to the
relation and, in  spite of the small scatter, it appears that, at given dynamical mass, galaxies
with more mass  in stars are younger than those with small stellar mass (see also
Fig.~\ref{lum_V}b). This weak trend cannot be accounted for entirely by the larger amount of mass
returned to the interstellar medium by evolved stars in older galaxies relative to younger ones.
For the \cite{chabrier03} IMF adopted here, the returned stellar mass fraction of a simple
stellar population increases by about 0.03~dex from $\log(t_r/{\rm yr})=9.5$ to 10 (with little
dependence on metallicity; the differential change is similar for a \citealt{salpeter} IMF). This
effect can thus account for only about 10  percent of the trend in $M_\ast/M_{dyn}$ with age in
Fig.~\ref{masses}.\footnote{For example, at $M_{dyn}\sim10^{11}M_\odot$, stellar mass increases
from $\log M_\ast/M_\odot\sim 10.75$ for $\log(t_r/{\rm yr})=10$ to $\sim 10.9$ for
$\log(t_r/{\rm yr})=9.5$.} The bulk of the trend might result from a systematically higher
baryonic fraction and/or higher efficiency of conversion of  baryons into stars in young
early-type galaxies relative to old ones. For example, if many of our early-types form by a
merger of star-forming systems, those which currently have the youngest  populations are
presumably the most recently merged and so spent the longest time in the star-forming phase.

We now examine in more detail how age, stellar metallicity and $\alpha$/Fe ratio depend on 
stellar and dynamical mass. This is shown in Fig.~\ref{par_sigma} for the early-type  galaxies in
our primary sample (Table~\ref{fits_par} provides simple linear fits to the relations shown in
the figure). The relations followed by age, metallicity and $\alpha$/Fe ratio as a function of
stellar mass (right-hand panels) reflect the conclusions drawn from  our analysis of the CMR and
$\mgsig$ relations in Section~\ref{cm} and \ref{mgv}. In  particular, we find that light-weighted
age increases (with a small scatter) with stellar mass in galaxies more massive than
$10^{11}M_\odot$, while there is a clear indication of a tail  towards younger ages in less
massive galaxies (Panel d; see also fig.12 of Paper~I). This confirms the results of several
previous studies of smaller samples of early-type galaxies at low- and slightly higher redshifts
\citep[e.g.][]{CR98,poggianti01b,thomas05,vandokkum03, treu05}. Stellar metallicity increases all
the way from the least massive to the most massive galaxies in our sample (Panel e). The relation
tends to steepen at stellar masses less than about $\sim3\times10^{10}M_\odot$, i.e. the
characteristic mass scale of several observed bimodalities (see Paper~I,
\citealt{kauf03b,baldry04}). We note that the dependence of stellar metallicity on stellar mass
found here for early-type galaxies resembles quite closely the relation presented by
\cite{christy04} for a sample of  SDSS star-forming galaxies.

\begin{figure*}
\centerline{\includegraphics[width=10truecm]{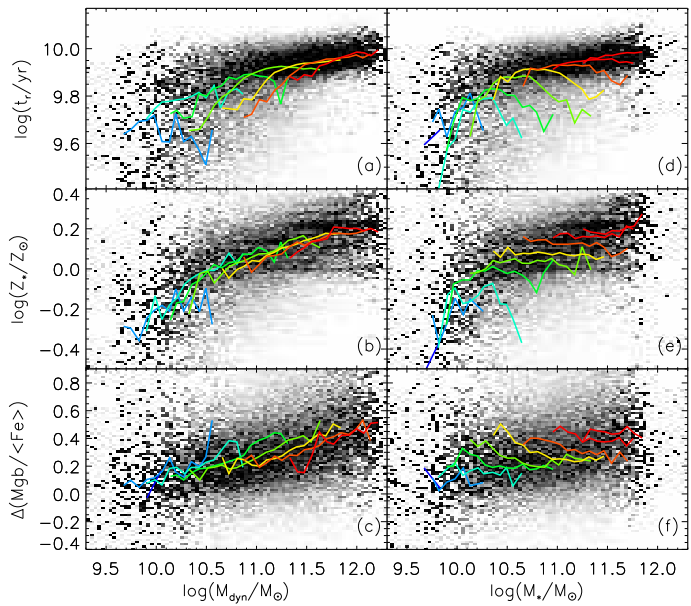}}
\caption{Left: Correlations between different physical parameters and the dynamical mass
estimated within the $r$-band Petrosian half-light radius (see equation~\ref{mdyn}): (a)
Light-weighted age; (b) stellar metallicity; (c) $\alpha$/Fe-estimator $\rm \Delta (\mgtfe)$. The
grey scale indicates the density of points at each location in the diagram, normalised to the
total number of galaxies  at fixed $\log(M_{dyn}/M_\odot)$. The solid lines represent the median
relations for galaxies in  different bins of stellar mass, centred on
$\log(M_\ast/M_\odot)=9.6,9.9,10.2,10.5,10.8,11.1,11.4,11.7$ (increasing from blue to red).
Right: Correlations between the same physical parameters and stellar mass. In this case, the
solid lines represent the median relations for galaxies in different bins of dynamical mass,
centred on $\log(M_{dyn}/M_\odot)=9.6,9.9,10.2,10.5,10.8,11.1,11.4,11.7,12$ (increasing from 
blue to red).}
\label{par_sigma}
\end{figure*}

The increase in metallicity with stellar mass, and hence the CMR, is a natural prediction of
different scenarios of early-type galaxy formation. In models in which  the galaxies form
monolithically in a single dissipationless collapse, star formation and chemical enrichment are
generally assumed to be interrupted by the onset of galactic winds following the major episode of
star formation. Since massive galaxies with deep potential wells are able to retain their gas for
much longer times and so to reach higher metallicities than  less massive galaxies, this scenario
accounts for the CMR as a metallicity sequence  \citep{larson74,bressan94,tantalo96,AY87}. Early
models of hierarchical galaxy formation neglected chemical enrichment and failed to produce
sufficiently red, luminous elliptical galaxies \citep{kauf93,cole94}. However, subsequent
renditions of these models have been able to reproduce the observed slope and scatter of the CMR
and of the $\mgsig$ relation by including chemical evolution and strong feedback even in massive
galaxies together with metallicity-dependent population synthesis models 
\citep{KC98,delucia04a}. In this respect, the CMR of early-type galaxies does not represent a
powerful tool to discriminate between the monolithic and merger formation scenarios
\citep{kaviraj05}. 

Additional clues about the formation of early-type galaxies may lie in their chemical abundance
patterns. Fig.~\ref{par_sigma}f shows that the $\alpha$/Fe abundance ratio, as traced by the
quantity $\rm \Delta(\mgtfe)$ (Section~\ref{sample}), increases linearly with  $\log M_\ast$.
This result, which was anticipated in several previous studies of early-type galaxies
\citep{wfg92,jorgensen99,greggio97}, has been quantified using stellar population models with
variable metal abundance ratios \citep{trager00a,mehlert03,thomas05}. If we  consider as
$\alpha$-enhanced those galaxies with $\rm \Delta(\mgtfe)>0.2$ (to account for the typical
observational error on \mgtfe), Fig.~\ref{par_sigma}f indicates that early-type galaxies with
solar abundance ratios are found only at $M_\ast \la 10^{10.5}M_\odot$,  corresponding roughly to
velocity dispersions $\sigma_V \la 100\,$km~s$^{-1}$. This is similar to the conclusion drawn by
\cite{kunt01}, based on the analysis of a sample of 72 early-type galaxies in groups and
clusters.

The observed range in $\alpha$/Fe abundance ratio and the increase of this ratio with stellar 
mass constitute a challenge for galaxy formation models. If the enrichment in $\alpha$ elements
relative to iron reflects the star formation timescale, as is assumed in standard chemical 
evolution models, the values $\rm \Delta(\mgtfe)>0.2$ found in high-mass early-type galaxies
require star formation timescales of the order 1--2 Gyr (see Section~\ref{summary}). Such short
timescales are plausible for the onset of galactic winds in the classical monolithic  collapse
scenario. Hierarchical models can also produce star formation histories which are  peaked at high
redshift for massive early-type galaxies, although subsequent star formation  is expected at
lower redshift in most models \citep{kauf96,thomas99}. A possible solution to this problem may be
the suppression of late gas cooling (and hence star formation) by  AGN-driven outflows in massive
early-type galaxies. As shown by various recent models of  hierarchical galaxy formation
accounting for the combined effects of star formation and black hole accretion, the feedback
provided by active galactic nuclei can halt star formation on  short timescales
\citep{granato01,granato04,springel05,croton05,delucia05}. This mechanism is expected to be most
effective in massive halos hosting supermassive black holes, for which the timescale to drive
outflows could be as short as 1~Gyr \citep{granato04}. Hence, the increase in the $\alpha$/Fe
abundance ratio with stellar mass adds important constraints to models of early-type galaxy
formation.

Fig.~\ref{par_sigma} also allows us to compare how age, metallicity and $\alpha$/Fe ratio vary
with dynamical mass (left-hand panels) versus stellar mass (right-hand panels). In the left-hand
panels, lines of different colours show the median relations between each parameter  and
$M_{dyn}$ followed by galaxies in different bins of stellar mass, increasing from 
$\log(M_\ast/M_\odot)=9.6$ to 11.7 in steps of $\sim$0.3~dex (from blue to red). In the 
right-hand panels, analogous lines show the median relations between each parameter and $M_\ast$
followed by galaxies in different bins of dynamical mass. Fig.~\ref{par_sigma}d shows that, at
fixed dynamical mass, galaxies with large $M_\ast$ tend to be younger than those with small 
$M_\ast$ (in agreement with the result of Fig.~\ref{lum_V}b above). This trend is noticeable  for
dynamical masses in the range $10^{10}\la M_{dyn}\la 10^{11.5}\,M_\odot$. In  contrast, at fixed
dynamical mass, stellar metallicity is almost independent of $M_\ast$  (Fig.~\ref{par_sigma}e).
In Fig.~\ref{par_sigma}f, galaxies with large $M_\ast$ tend to have slightly lower $\rm
\Delta(\mgtfe)$ and hence $\alpha$/Fe than those with small $M_\ast$ at fixed $M_{dyn}$. This is
consistent with the trend between $\rm \Delta(\mgtfe)$ and  age identified in Fig.~\ref{cube}
below. A remarkable conclusion from the comparison between the right- and left-hand panels in
Fig.~\ref{par_sigma} is that the chemical composition of early-type galaxies appears to depend
primarily on dynamical mass rather than stellar mass. 

\begin{figure*}
\centerline{\includegraphics[width=12truecm]{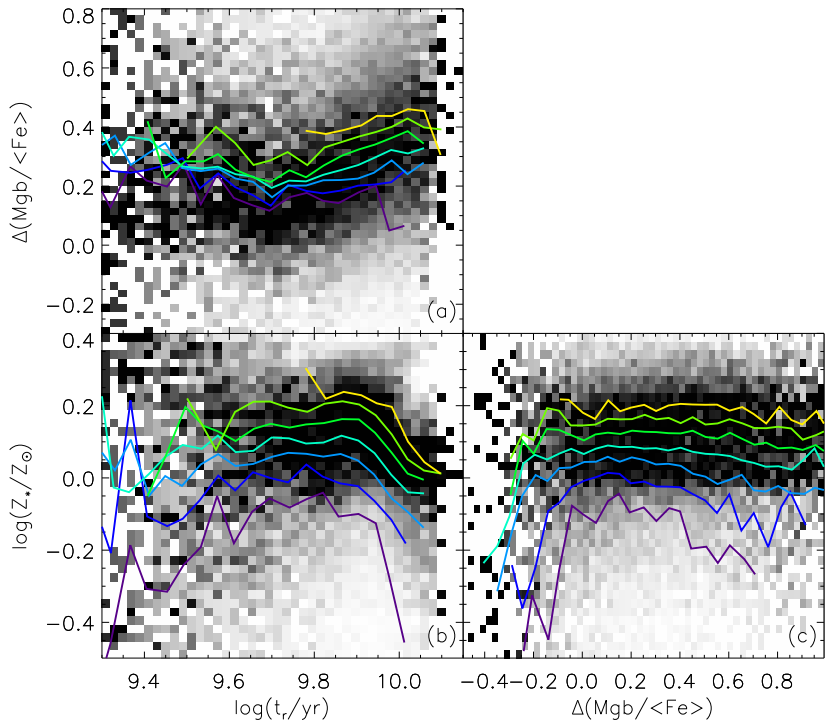}}
\caption{Correlations between stellar metallicity, light-weighted age and $\alpha$/Fe-estimator
$\rm \Delta(\mgtfe)$. The solid lines represent the median relations for galaxies in different
bins of velocity dispersion, centred on $\log(\sigv/\rm km\,s^{-1})=1.9,2,2.1,2.2,2.3,2.4,2.5$
(increasing from purple to yellow). The grey scale indicates the density of galaxies with respect
to the total at fixed abscissa.}\label{cube}
\end{figure*}

It is also of interest to examine the relations between age, metallicity and $\alpha$/Fe ratio.
These are shown in Fig.~\ref{cube} for our primary sample of early-type galaxies. In each panel,
lines of different colours indicate the median trends followed by galaxies in  different bins of
velocity dispersion, from $\sigma_V\sim80$~km~s$^{-1}$ (purple) to  $\sim300$~km~s$^{-1}$
(yellow). Fig.~\ref{cube}b shows that, at fixed velocity dispersion,  age and metallicity are
strongly anticorrelated for galaxies older than $\log (t_r/{\rm yr}) =9.8$. \cite{bernardi05},
who {\em assume} that the scatter about the CMR is caused entirely by  age variations, suggest
that young galaxies may be more metal-rich than older galaxies at fixed velocity dispersion.
Other previous studies also invoked an age-metallicity  anticorrelation to explain the tightness
of the colour-magnitude and the $\mgsig$  relations (e.g.,
\citealt{jorgensen99,fcs99,trager00b,tf01}, but see \citealt{kunt01}). However, the slope of this
anticorrelation in Fig.~\ref{cube}b is consistent with the slope of the  age-metallicity
degeneracy for individual galaxies quantified in Paper~I using the same sample of early-type
galaxies. Thus, we cannot conclude here whether this anticorrelation is real or induced by
correlated errors, but we note that the apparent scatter in age of massive, old ellipticals is
consistent with the measurement error and the scatter in metallicity is quite small.

Fig.~\ref{cube}c shows the relation between $\rm \Delta(\mgtfe)$ and metallicity for the galaxies
in our sample. Interestingly, even though both the colour-magnitude and the  $\mgsig$ relations
imply an increase in both total metallicity and $\alpha$/Fe  with mass, we do not identify any
significant correlation between these two quantities (but see Fig.~\ref{alpha}a below). This is
true for the sample as a whole and for subsets of galaxies with fixed velocity dispersion.
Instead, $\rm \Delta(\mgtfe)$ appears to  correlate more strongly with light-weighted age for
$\log (t_r/{\rm yr}) \ga 9.7$, independent of galaxy velocity dispersion (Fig.~\ref{cube}a). Old
galaxies appear to have larger  $\alpha$/Fe ratio than younger ones. This may be expected from
the late enrichment in iron by Type~Ia supernovae in galaxies which completed their star
formation more recently, if  the $\alpha$/Fe ratio reflects the star formation timescale. We
might be tempted to relate star formation efficiency (and hence $\alpha$-enhancement) to stellar
surface mass density, if the scaling between star formation rate and gas surface mass density in
spiral galaxies can be extended to the progenitors of early-type galaxies
\citep[e.g.][]{schmidt59,kennicutt98}. Fig.~\ref{alpha}b shows $\rm \Delta(\mgtfe)$ as a function
of surface stellar mass  density for the galaxies in our sample. The lines of different colours
indicate the median trends followed by galaxies in different age bins, from $t_r\sim3$ to 10 Gyr.
We do not detect any significant correlation between $\rm \Delta(\mgtfe)$ and surface stellar
mass density. For completeness, in Fig.~\ref{alpha}a, we show again $\rm \Delta(\mgtfe)$ as  a
function of metallicity, as in Fig.~\ref{cube}c (the grey scale in this case indicates the
distribution of galaxies as a function of $\rm \Delta(\mgtfe)$ at fixed metallicity). There
appears to be a mild correlation between $\alpha$/Fe and metallicity for galaxies in fixed  age
bins, in agreement with the results of \cite{trager00a}. 

\begin{figure}
\centerline{\includegraphics[width=10truecm]{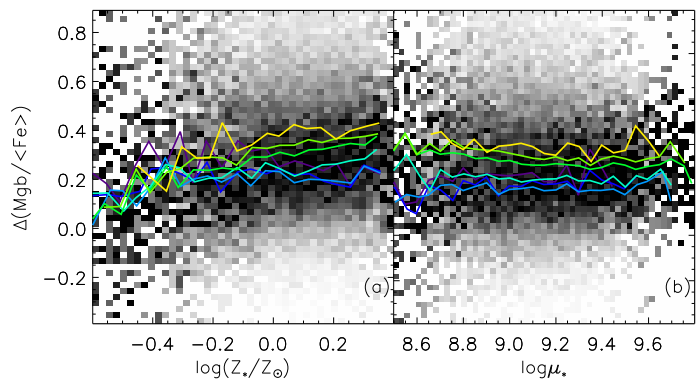}}
\caption{$\alpha$/Fe-estimator $\rm \Delta(\mgtfe)$ plotted against (a) stellar metallicity and
(b) surface stellar mass density within the $r$-band Petrosian half-light radius \rpetro. The
solid lines represent the median relations for galaxies in different bins of light-weighted age,
centred on $\log(t_r/{\rm yr})=9.5,9.6,9.7,9.8,9.9,10,10.1$ (increasing from purple to yellow).}
\label{alpha}
\end{figure}

\section{Summary and conclusions}\label{summary}
We have analysed a sample of 26,003 early-type galaxies selected from the SDSS DR2 on the basis
of their light concentration. Light-weighted ages, stellar metallicities and stellar masses for
this sample were previously derived through the comparison of a set of carefully selected
spectral absorption features with a comprehensive library of  high-resolution population
synthesis models, encompassing the full range of physically plausible star formation histories
(paper~I). In addition to these physical parameters, we have considered here an empirical
estimate of the $\alpha$-elements-to-iron abundance ratio, given by the offset $\rm
\Delta(\mgtfe)$ between the observed ratio of the Mgb and $\rm \langle Fe\rangle$ indices of a
galaxy and that predicted by the best fitting model in the library  (which has scaled-solar
abundance ratios). We have used these data to investigate  the physical origin of two well-known
scaling relations for early-type galaxies: the colour-magnitude and the $\mgsig$ relations.

Our analysis demonstrates unambiguously and with unprecedentedly good statistics that both the 
colour-magnitude and the $\mgsig$ relations are primarily sequences in galaxy stellar mass. At
increasing stellar mass, as traced by either luminosity or velocity  dispersion, the increasing
colour and Mg-absorption line strength along the relations reflect an increase in both total
metallicity and $\alpha$/Fe ratio. Moreover, the galaxies in our sample cover a range in age of
about $3-4$ Gyrs, with more massive galaxies being on average older than low-mass galaxies. While
at high masses early-type galaxies have the same mean age (and a small scatter in metallicity),
at lower masses there is an increasing spread toward younger ages. This age spread dominates the
scatter about the observed relations at low masses, in the sense that younger  galaxies deviate
toward bluer colours and lower \mgtwo\ index strengths than older galaxies of  the same mass.
These results are consistent with the conclusions from previous studies based on smaller samples
of early-type galaxies \citep{kodama97,colless99,vazdekis01a, worthey03}. In addition, we find
that the scatter in metallicity at fixed stellar mass contributes significantly to the scatter
about the two observational scaling relations, in particular at high masses.

We have checked that our main conclusions are not affected by possible dust effects and that they
are robust against sample selection. In particular, the possible contamination of our sample by
bulge-dominated star-forming galaxies, which could amount to $\sim$10  percent, does not
substantially affect on our results.

For a small subsample of 1765 galaxies we used information on environmental density available
from \cite{kauf04} to  explore the dependence of the observed scaling relations and the stellar
physical parameters on environment. We have found a small but detectable difference in the
zero-point of the two relations, in the sense that early-type galaxies in dense environments tend
to have redder colours and stronger $\rm Mg_2$ absorption indices than galaxies in low-density
regions, at fixed luminosity or velocity dispersion. We also find a systematic increase in the
scatter about both relations from high to low densities. These variations appear to be induced by
small differences in the light-weighted age and metallicity of galaxies located in different
environments. While galaxies with similar mass have the same element abundance ratio regardless
of environment, there is an increasing spread toward younger ages and lower metallicities in
low-density environments. At fixed stellar mass, early-type galaxies in dense regions are on
average 0.02~dex older and more metal-rich than early-type galaxies in low-density regions. We
note that these trends are very small and it will be worth re-examining them when a better
statistics is available. If confirmed, these results are in agreement with previous studies
indicating that early-type galaxies in clusters started to form stars earlier than, but on the
same timescale as early-type galaxies in the field \citep[e.g.][]{thomas05,bernardi06,clemens06}.

We have also studied the dependence of the stellar mass on the dynamical mass estimated within
the $r$-band Petrosian half-light radius of a galaxy. The relation is well described by a power
law of exponent $0.783\pm0.019$ (equation~\ref{masseq}), implying a decrease in the
stellar-to-dynamical mass ratio from low- to high-mass galaxies. The correlations of physical
parameters with the dynamical mass estimated in this way suggests that metallicity and  element
abundance ratios in early-type galaxies are more fundamentally related to dynamical mass than to
stellar mass.

The increase in total metallicity with dynamical mass favours the classical interpretation of the
colour-magnitude and \mgsig\ relations in terms of supernova-driven winds \citep[e.g.]
[]{trager00b,thomas05}. To account for the simultaneous increase in total metallicity and
$\alpha$/Fe ratio with mass, galactic winds should occur at early times, i.e. prior to the onset
of Type~Ia supernovae (on a timescale of a few Gyr), which are the main contributors to Fe-peak
elements. The winds will be more effective in removing $\alpha$ elements (produced by Type~II
supernovae on a time scale of $\sim10^8\,$yr) from low-mass galaxies with shallow  potential
wells, while they should not reduce significantly the fraction of primordial gas and hence star
formation \citep{maclow99}. Much observational evidence has been accumulated for the importance
of galactic outflows in galaxies with masses up to at least $10^{10}M_\odot$
\citep{LH96,heckman00,pettini00}. Our results, if interpreted in terms of galactic winds,
indicate that even more massive galaxies (with masses up to $\sim10^{11}M_\odot$ in stars) have
been affected by the ejection of metals through galactic winds.

The above scenario, however, cannot account alone for the observed values of $\alpha$/Fe in
early-type galaxies. In a galactic-wind scenario, massive galaxies are predicted to have  solar
$\alpha$/Fe abundance ratios, while low-mass galaxies, which lose $\alpha$ elements at  early
times, should have lower than solar $\alpha$/Fe ratios. Instead, galaxies with stellar masses
less than about $3\times10^{10}M_\odot$ (corresponding to velocity dispersions less than
$\sim100$~km~s$^{-1}$) are observed to have nearly solar $\alpha$/Fe ratios, while this ratio
increases to super-solar values in more massive galaxies (the quantity $\rm \Delta(\mgtfe)$
reaching values around 0.3 in galaxies with stellar masses near $3\times10^{11}M_\odot$; see 
Fig.~\ref{par_sigma}).

The super-solar $\alpha$/Fe abundance ratios of massive early-type galaxies suggest that these
formed on a relatively short timescale and/or have/have had an IMF skewed towards high-mass 
stars. An IMF enriched in massive stars will produce a larger ratio of Type~II to Type~Ia 
supernovae, and hence a larger $\alpha$/Fe ratio. \cite{nagashima05} have shown that a top-heavy
IMF during the burst ignited by the major merger that formed an elliptical galaxy can reproduce
the observed range in $\alpha$/Fe ratios. However, none of the models they explore yields the
observed correlation of $\alpha$/Fe with velocity dispersion (a model in which thermal conduction
prevents the gas from cooling at the centres of massive halos is able to produce  an increase in
the $\alpha$-element abundance, but not the $\alpha$/Fe ratio, with mass).

An interpretation of the $\alpha$/Fe ratio in terms of the star formation timescale is supported
by the correlation we find between $\rm \Delta(\mgtfe)$ and light-weighted age, independent of
mass (Fig.~\ref{cube}). This suggests that galaxies with longer star formation timescales (and
thus with more recently formed stellar populations) have lower $\alpha$/Fe ratio than galaxies
formed on shorter timescales, because they had time to recycle the Fe-peak elements ejected by
Type~Ia supernovae. We also find that light-weighted age increases with stellar mass with
negligible scatter at masses above $10^{11}M_\odot$ (Fig.~\ref{par_sigma}d). Early-type galaxies
less massive than about  $\sim10^{11}M_\odot$ display an extended tail toward younger ages, the
mean age declining markedly with decreasing mass. This  suggests either that low-mass galaxies
formed more recently than high-mass galaxies, or that they have a more extended star formation
history (consistent with their solar $\alpha$/Fe ratios). 

Our results represent further evidence for a shift in stellar growth toward less massive galaxies
in recent epochs \citep{cowie96,delucia04b,kodama04,Yi05,treu05}. This `downsizing' scenario may
appear at odds with the expectations of original hierarchical models of galaxy formation.
However, observations and the hierarchical paradigm can be reconciled if detailed physics of
feedback from supernovae, active galactic nuclei or thermal conduction is introduced
\citep[e.g.][]{benson03,granato04,nagashima05a}.  These sources of feedback could inhibit star
formation on timescales short enough for the bulk of the star formation to be completed before
Type~Ia supernovae can substantially increase the iron abundance in massive galaxies.
\cite{springel05} have shown that, in major mergers of spiral galaxies hosting supermassive 
black holes, AGN feedback provides a mechanism that can quench star formation on short 
timescales. This mechanism is more efficient in the most massive early-type galaxies and leaves
dwarf spheroids almost unaffected. We also note that the short star formation  timescales (i.e.
high formation redshifts) of massive early-type galaxies do not preclude longer assembly
timescales \citep{delucia05}: massive early-type galaxies  could appear old even if they
assembled relatively recently. In this context, the new  constraints derived here on the physical
origin of the colour-magnitude and \mgsig\ relations for early-type galaxies represent a valuable
reference for future models.

\section*{Acknowledgments}

We thank Mariangela Bernardi for helpful discussion on sample selection and the anonymous referee
for valuable suggestions that have improved our analysis. A.G. and S.C. thank the Alexander von
Humboldt Foundation, the Federal Ministry of Education and Research, and the Programme for
Investment in the Future (ZIP)  of the German Government for funding through a Sofja Kovalevskaja
award. A.G. thanks the European Association for Research in Astronomy training  site (EARA) and
the European Community for a Marie Curie EST fellowship  (MEST-CT-2004-504604). J.B. acknowledges
the receipt of FCT fellowship BPD/14398/2003.

Funding for the creation and distribution of the SDSS Archive has been provided by the Alfred P.
Sloan Foundation, the Participating Institutions, the National Aeronautics and Space
Administration, the National Science Foundation, the US Department of Energy, the Japanese
Monbukagakusho, and the Max Planck Society. The SDSS Web site is http://www.sdss.org/. The
Participating Institutions are the University of Chicago, Fermilab, the Institute for Advanced
Study, the Japan Participation Group, the Johns Hopkins University, the Max Planck Institute for
Astronomy (MPIA), the Max Planck Institute for Astrophysics (MPA), New Mexico State University,
Princeton University, the United States Naval Observatory, and the University of Washington.

\label{lastpage}
\end{document}